\documentclass[journal]{IEEEtran}

\ifCLASSINFOpdf
\usepackage[pdftex]{graphicx}
\DeclareGraphicsExtensions{.pdf,.jpeg,.png}
\else
\usepackage[dvips]{graphicx}
\fi
\usepackage{subfigure}
\usepackage{epstopdf}
\usepackage[cmex10]{amsmath}
\usepackage{amssymb}
\usepackage{wasysym}
\usepackage{algorithm}
\usepackage{algorithmic}
\usepackage{array}
\usepackage{cite}
\usepackage{color}
\usepackage{url}
\usepackage{bm}
\usepackage{enumerate}
\usepackage{epsfig,latexsym}
\usepackage{cite}
\begin{document}
%
\title{Beam Training and Tracking for Extremely Large-Scale MIMO Communications}

\author{Kangjian~Chen,~\IEEEmembership{Student~Member,~IEEE}, Chenhao~Qi,~\IEEEmembership{Senior~Member,~IEEE},\\  Cheng-Xiang Wang,~\IEEEmembership{Fellow,~IEEE}, and Geoffrey Ye Li,~\IEEEmembership{Fellow,~IEEE}
\thanks{This work was supported in part by the National Natural Science Foundation of China under Grants 62071116, U22B2007 and 61960206006, in part by the Fundamental Research Funds for the Central Universities under Grant 2242022k60006, in part by the Key Technologies R\&D Program of Jiangsu (Prospective and Key Technologies for Industry) under Grants BE2022067 and BE2022067-1, in part by the EU H2020 RISE TESTBED2 project under Grant 872172, and in part by the Postgraduate Research \& Practice Innovation Program of Jiangsu Province under Grant KYCX23\_0262. This article has been presented in part at the 2022 IEEE/CIC International Conference on Communications in China (ICCC), Foshan, China, Aug. 2022~\cite{ICCC2022CKJ}. (\textit{Corresponding author: Chenhao~Qi})}
	\thanks{Kangjian~Chen and Chenhao~Qi are with the School of Information Science and Engineering, Southeast University, Nanjing 210096, China (e-mail: qch@seu.edu.cn).}
	\thanks{Cheng-Xiang Wang is with the National Mobile Communications Research Laboratory, School of Information Science and Engineering, Southeast University, Nanjing 210096, China, and also with the Purple Mountain Laboratories, Nanjing 211111, China (e-mail: chxwang@seu.edu.cn).}
	\thanks{Geoffrey Ye Li is with the Department of Electrical and Electronic Engineering, Imperial College London, SW7 2AZ London, U.K. (e-mail: geoffrey.li@imperial.ac.uk).}
}

\markboth{Accepted by IEEE Transactions on Wireless Communications}
{}

\maketitle

\begin{abstract}
In this paper, beam training and beam tracking are investigated for extremely large-scale multiple-input-multiple-output communication systems  with partially-connected hybrid combining structures. Firstly, we propose a two-stage hybrid-field beam training scheme for both the near field and the far field. In the first stage, each subarray independently uses multiple far-field channel steering vectors to approximate near-field ones for analog combining. To find the codeword best fitting for the channel, digital combiners in the second stage are designed to combine the outputs of the analog combiners from the first stage.  Then, based on the principle of stationary phase and the time-frequency duality, the expressions of subarray signals after analog combining are analytically derived and a beam refinement based on phase shifts of subarrays~(BRPSS) scheme with closed-form solutions is proposed for high-resolution channel parameter estimation. Moreover, a low-complexity near-field beam tracking scheme is developed, where the kinematic model is adopted to characterize the channel variations and the extended Kalman filter is exploited for beam tracking. Simulation results verify the effectiveness of the proposed schemes.
\end{abstract}


\section{Introduction}

Massive multiple-input-multiple-output (MIMO) is a key technology for the fifth-generation wireless communications~\cite{ICCC2022CKJ,WC21CZ}. Equipped with large antenna arrays, the base station (BS) can achieve high spectral efficiency dramatically by exploiting the spatial degree of freedom~\cite{TWC20QCH,OPCS23NBY}.  To further enhance the spectral efficiency for the future sixth-generation wireless communications, extremely large-scale MIMO (XL-MIMO) that uses far more antennas than the existing massive MIMO is developed~\cite{CST23_WCX,SciChinaXiaohu2020,JSAC23WZD}.

Due to high power consumption, the fully-digital structure that allocates each antenna with a dedicated radio frequency (RF) chain is impractical for large antenna arrays~\cite{TWC20SXS}. Consequently, the hybrid precoding structure, where a small number of RF chains are connected to a large number of antennas, is developed for XL-MIMO systems~\cite{Tcom22CMH}. According to the ways how the RF chains are connected to the antennas, the hybrid precoding structure can generally be divided into two categories, i.e., fully-connected hybrid precoding structure and partially-connected hybrid precoding structure. Although the fully-connected hybrid precoding structure can achieve better spectral efficiency than the partially-connected hybrid precoding structure, the latter is more practical than the former, owing to its low hardware complexity as well as its flexibility to be extended to different sizes of antennas in blocks~\cite{TWC20NS}.

One important difference between XL-MIMO and the existing massive MIMO is the channel features. According to the distance between the user and the BS, the radiation field can be divided into the near field and the far field. The boundary to separate the near field and the far field is referred to as the Rayleigh distance~\cite{TWC22LHQ,APM17SKT,TWC23SX}. On one hand, the Rayleigh distance increases linearly with the wavelength. On the other hand, when fixing the wavelength, the Rayleigh distance increases quadratically with the number of antennas. As a result, the far field assumption in existing massive MIMO may not hold for the XL-MIMO, especially when a user is close to the BS. In this context, channel state information (CSI) acquisition methods for XL-MIMO should consider both the near-field and far-field effects.  

In general, CSI acquisition includes channel estimation and beam training~\cite{SciChinaChenhao2021}. Channel estimation usually focuses on efficient estimation of the high-dimensional channel matrix by exploiting advanced signal processing techniques, such as compressed sensing. However, the beam training can avoid the estimation of the high-dimensional channel matrix and obtain considerable beamforming gain, especially in low signal-to-noise ratio (SNR). In~\cite{Tcom22CMH}, to estimate the near-field channels in XL-MIMO, a polar-domain simultaneous orthogonal matching pursuit (P-SOMP) algorithm is proposed, where random beamforming instead of the directional beamforming is used. For the two-phase beam training (TPBT) method in~\cite{WCL22ZYP}, the first phase determines the candidate channel angles while the second phase finds the channel distance based on shortlisted candidate angles from the first phase. For the distance-based hierarchical beam training (DHBT) method in~\cite{CC22WXH}, the codebook is designed by equally sampling the distance. Although these methods work well for the fully-connected hybrid precoding structure, they are not tailored for the practical partially-connected  hybrid precoding structure.

One common challenge in beam training is the limited resolution of the predefined codebook. To achieve high-resolution channel parameter estimation, the low-complexity beam refinement is widely adopted~\cite{TVT21SH,TVT23ZA,TWC17ZDL}.  In~\cite{TVT21SH}, based on the monopulse signals, beam refinement with a closed-form solution is developed. In~\cite{TVT23ZA}, an efficient angle-of-arrival estimator is designed by approximating the power of the array response as a Gaussian function.   In~\cite{TWC17ZDL}, an auxiliary beam pair is designed to provide high-resolution estimates for the angles of the channel. These beam refinement methods are for far-field channels and more investigations are desired for near-field channels.

With far more antennas than massive MIMO, XL-MIMO systems suffer from the heavy burden of training overheads. An efficient way to reduce the training overhead is the beam tracking, which exploits the correlations of the channels at different time instants to narrow down the sets of training candidates~\cite{Arxiv23YCS}. To meet the demands for high reliability and massive connectivity in future wireless communications, beam tracking is usually characterized by high beamforming gain, low training overhead and low computational complexity. There are a variety of beam tracking schemes for the far-field channels~\cite{TWC18ZDL,JSAC21TJB,TWC21LCS,TWC22NBY}. In~\cite{TWC18ZDL}, well-designed pairs of auxiliary beams can capture the angle variations. The beam zooming-based beam tracking scheme  in~\cite{JSAC21TJB} exploits the delay-phase precoding structure to flexibly control the angular coverage of frequency-dependent beams. In the  adaptive tracking framework in \cite{TWC21LCS}, beam direction is updated according to measurements of the current data beam. The grid-based hybrid tracking scheme in~\cite{TWC22NBY} searches the surroundings of the former beam and selects the best beam according to the changing trend of the previously used beams.   For the near field, user tracking has also been investigated in a variety of works, such as \cite{JSTSP19ZWL,TSP21GA,Arxiv23PS}. However, the approaches of these works are incompatible with the beam training framework. More investigations are needed for efficient beam tracking in the near field.

In this paper, for XL-MIMO systems with  partially-connected hybrid combining structure, we investigate beam training, beam refinement and beam tracking, which are three progressive techniques to accomplish the beam alignment. Our contributions are summarized as follows.

\begin{itemize}
\item We propose a two-stage hybrid-field beam training (THBT) scheme, which works for both the near and far fields. In the first stage, we use far-field channel steering vectors of  subarrays to approximate the near-field ones for analog combining so that beam training for both the near and far fields can be performed simultaneously. In the second stage, digital combiners are designed to combine the outputs of the analog combiners from the first stage. Then from the predefined hybrid-field codebook, we select the codeword corresponding to the dedicated digital combiner that achieves the largest combining power as the result of the THBT.
	
\item We propose a beam refinement based on phase shifts of subarrays (BRPSS) scheme. Based on the principle of stationary phase (PSP) and the time-frequency duality, the expressions of subarray signals after analog combining are analytically derived, where the phases of these signals change quadratically with the subarray indices. By exploiting the phase shifts of subarrays, closed-form estimates of channel parameters can be obtained. 

\item We develop a low-complexity near-field beam tracking (NFBT) scheme. As the near-field channels are related to both the angle and distance of the radiation source, the kinematic model is adopted to characterize the channel variations. In addition, the BRPSS scheme is used to estimate the real-time channel parameters. Then the kinematic model and real-time estimates are exploited by the extended Kalman filter (EKF) to track and predict the near-field channel parameters.
\end{itemize}

The rest of this paper is organized as follows. Section~\ref{SystemModel} introduces the model of the XL-MIMO systems. In Section~\ref{SecTSHFBT}, we propose the two-stage hybrid-field beam training scheme. The beam refinement based on phase shifts of subarrays scheme is provided in Section~\ref{SecBRPSS}. The near-field beam tracking  scheme is discussed in Section~\ref{NFBT}. The simulation results are presented in Section~\ref{SimulationResults}, and the paper is concluded in Section~\ref{Conclusion}.

The notations are defined as follows. Symbols for matrices (upper case) and vectors (lower case) are in boldface. The set is represented by bold Greek letters. $(\cdot)^{\rm T}$ and $(\cdot)^{\rm H} $ denote the transpose and conjugate transpose (Hermitian), respectively. $\left[ \boldsymbol{A} \right] _{:,m}$ denotes the $m$th column of a matrix $\boldsymbol{A}$. $j$ denotes the square root of $-1$. In addition, $|\cdot |$ and $\|\cdot \|_2$ denote the absolute value of a scalar and $\ell_2$-norm of a vector, respectively. $\mathbb{C}$ denotes the set of complex numbers. The complex Gaussian distribution is denoted by $\mathcal{C}\mathcal{N}$. $\lfloor\cdot\rfloor$ and ${\rm blkdiag}\{\cdot\}$ denote the floor operation and the block diagonalization operation, respectively. Moreover, $\mathcal{O}$ and $\mbox{mod}(\cdot)$ denote the order of complexity and the operation of modulo, respectively.

\section{System Model}\label{SystemModel}
\begin{figure}[!t]
	\centering
	\includegraphics[width=66mm]{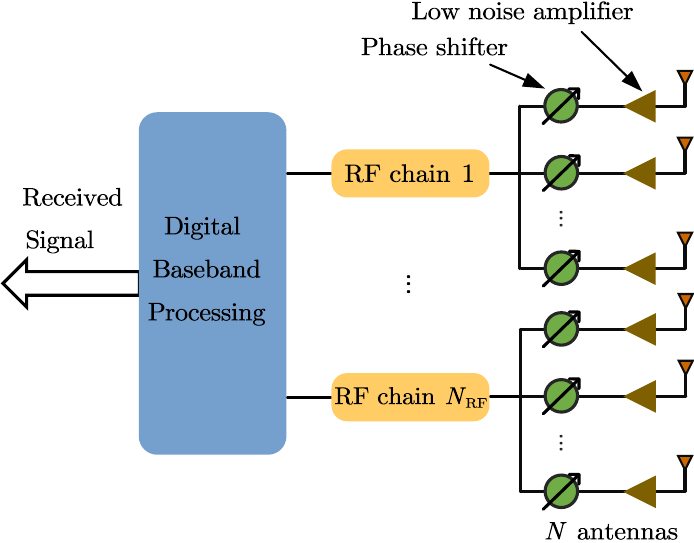} 
	\caption{Illustration of a partially-connected hybrid combining structure.}
	\label{fig:system model}
	\vspace{-0.4cm}
\end{figure}

We consider uplink beam training for an XL-MIMO system with a user and a BS. As shown in Fig.~\ref{fig:system model}, the BS employs a large-scale uniform linear array (ULA) of $N$ antennas with half wavelength interval and a partially-connected hybrid combining structure with $N_{\rm RF}$ RF chains, where the hybrid combining includes analog and digital combining. In practice, the BS with a hybrid structure has far more antennas than RF chains, i.e., $ N \gg N_{\rm RF}$. The ULA is formed by $N_{\rm RF}$ non-overlapping subarrays, where each subarray has $M = N/N_{\rm RF}$ antennas and is solely connected to an RF chain after analog combining. Then all the $N_{\rm RF}$ RF chains are connected to a digital baseband processing unit for digital combining. In this work, we focus on the processing at the BS and consider a single-antenna user for simplicity. However, the proposed schemes can be extended to scenarios with multi-antenna~users.


During uplink beam training, the $p$th transmit pilot signal by the user is denoted as $x_p$ for $p=1,2,\ldots,P$, where $P$ is the pilot length. The channel vector between the user and the BS is denoted as $\boldsymbol{h}\in\mathbb{C}^N$. Then the signal after the analog combining at the BS side can be expressed as~\cite{TWC20NS}
\begin{equation}\label{system model}
{y}_p = \boldsymbol{v}_p \boldsymbol{W}_p \boldsymbol{h} x_p + \boldsymbol{v}_p \boldsymbol{W}_p \boldsymbol{\eta},
\end{equation}
where $\boldsymbol{W}_p\in \mathbb{C}^{N_{\rm RF} \times  N}$ is the analog combiner, $\boldsymbol{v}_p\in \mathbb{C}^{1 \times  N_{\rm RF}}$ is the digital combiner, and $\boldsymbol{\eta}\in \mathbb{C}^{N}$ is an additive white Gaussian noise (AWGN) vector obeying $\boldsymbol{\eta}\sim\mathcal{C}\mathcal{N}\left( \boldsymbol{0},\sigma ^2 \boldsymbol{I}_{N} \right) $. If the digital combiner is a matrix, we may treat each row of it as a different $\boldsymbol{v}_p$, so that we can perform parallel baseband processing.

Generally, two kinds of channels, namely  the near-field channel and the far-field channel, are considered in the existing literature according to the distance between the user and the BS~\cite{TWC22LHQ,APM17SKT}. The commonly used boundary distance to distinguish the near field and the far field is the Rayleigh distance 
\begin{align}\label{Rayleigh distance}
	Z = 2D^2 /\lambda,
\end{align}
where $D$ denotes the array aperture and $\lambda$ denotes the wavelength. In other words, when the distance between the user and the BS is larger than $Z$, the wireless channel is regarded as the far-field channel; otherwise, it is the near-field channel. Since a half-wavelength-interval ULA is adopted in this work, the array aperture at the BS is $D = N\lambda/2$, and therefore $Z = N^2\lambda/2$.

\begin{figure}[!t]
	\centering
	\includegraphics[width=67.35mm]{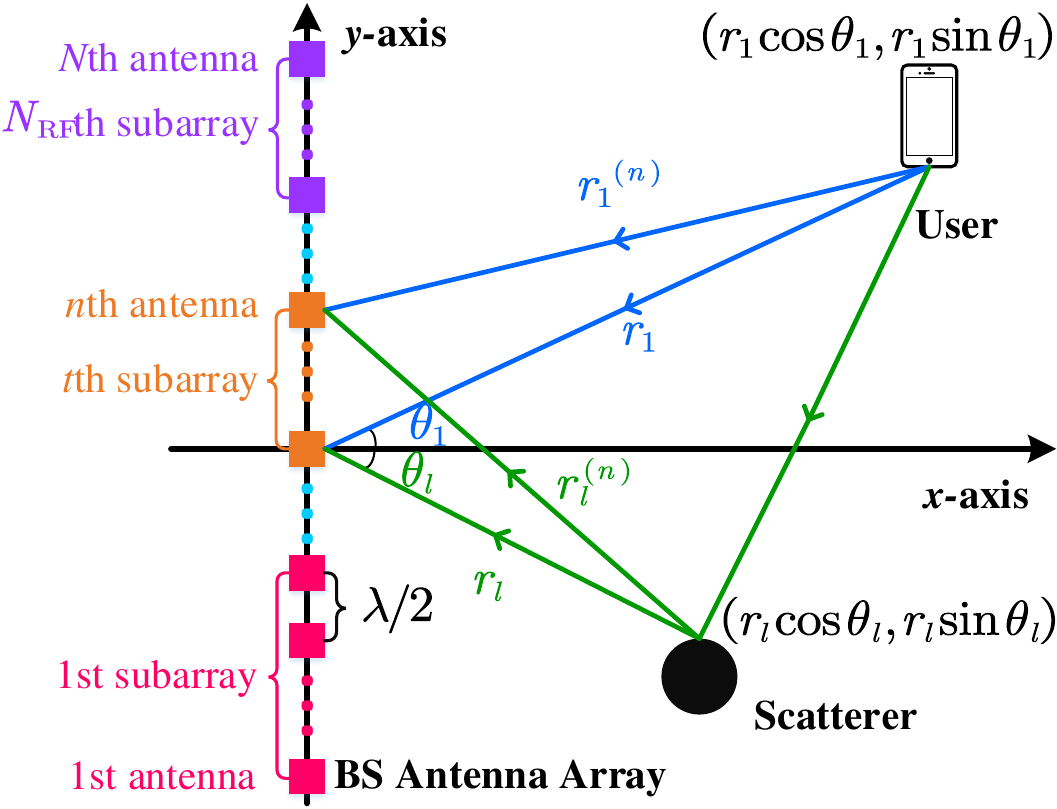}
	\caption{Illustration of a near-field channel model.}
	\label{MultipathChannelModel}
		\vspace{-0.265cm}
\end{figure}

As shown in Fig.~\ref{MultipathChannelModel}, a near-field channel composed of one line-of-sight (LoS) path  and several non-line-of-sight (NLoS) paths between the BS and the user is considered. $N$ antennas of the BS are placed along the y-axis in the Cartesian coordinate system and the coordinate of the $n$th antenna is $(0,\delta_n\lambda)$, where $\delta_n\triangleq(2n-N-1)/4$ for $n=1,2,\ldots,N$. The coordinate of the user is denoted as $\boldsymbol{p}_1 = (r_1\cos\theta_1,r_1\sin\theta_1)$, where $r_1$ is the distance between the user and the origin, and $\theta_1\in[-\pi/2,\pi/2]$ is the angle of the user relative to the positive x-axis. Similarly, the coordinate of the scatterer on the $l$th path for $l\ge2$ is denoted as $\boldsymbol{p}_l=(r_l\cos\theta_l,r_l\sin\theta_l)$, where $r_l$ is the distance between the scatterer on the $l$th path and the coordinate origin, and $\theta_l\in[-\pi/2,\pi/2]$ is the angle of the  scatterer on the $l$th path relative to the positive x-axis. The distance between $\boldsymbol{p}_l$  and the $n$th antenna can be expressed as
\begin{equation}\label{Eq.RangeRln}
r_l^{(n)} = \sqrt{r_l^2 + \delta_n^2\lambda^2 - 2r_l\Omega_l\delta_n\lambda},
\end{equation}
for $l\ge 1$, where $\Omega_l\triangleq\sin\theta_l \in[-1,1]$. Then the channel vector between the user and the BS can be expressed as
\begin{equation}\label{Eq.channel model}
\boldsymbol{h} = \sum_{l=1}^{L}g_l\boldsymbol{\alpha}(N,\Omega_l,r_l),
\end{equation}
where $L$ and $g_l$ denote the number of paths and the path gain of the $l$th path, respectively. The channel steering vector, $\boldsymbol{\alpha}(\cdot)$, is defined as
\begin{equation}\label{Eq.steering vector}
\boldsymbol{\alpha}(N,\Omega_l,r_l)\! =\! \frac{1}{\sqrt{N}}\!\left[e^{-j\frac{2\pi}{\lambda}(r_l^{(1)}\!-\!r_l)},\!\ldots\!,e^{-j\frac{2\pi}{\lambda} (r_l^{(N)}-r_l)}\right]^{\rm T}.
\end{equation}
Note that the channel steering vector in \eqref{Eq.steering vector} can be used to describe both the far-field channel and the near-field channel. 

If $r_l>Z$, $r_l^{(n)}$ in \eqref{Eq.RangeRln} can be approximated  as $r_l^{(n)} \approx r_l - \Omega_l\delta_n\lambda$ because $\delta_n\lambda/r_l\approx0$ and $\sqrt{1+\epsilon}\approx1+\frac{1}{2}\epsilon$~\cite{Tcom22CMH}. As a result, $\boldsymbol{\alpha}(N,\Omega_l,r_l)$ in \eqref{Eq.steering vector} can be approximated as
\begin{align} \label{FFSteeringVector}
	\boldsymbol{\alpha}(N,\Omega_l,r_l) &\approx \frac{1}{\sqrt{N}}\left[e^{j2\pi\Omega_l\delta_1},\ldots,e^{j2\pi\Omega_l\delta_N}\right]^{\rm T}.
\end{align}
Multiplying a constant phase factor $e^{-j2\pi\Omega_l\delta_1}$ to all the entries of $\boldsymbol{\alpha}(N,\Omega_l,r_l)$ in \eqref{FFSteeringVector}, we have
\begin{align}
e^{-j2\pi\Omega_l\delta_1}\boldsymbol{\alpha}(N,\Omega_l,r_l) &\approx \frac{1}{\sqrt{N}}\left[1,e^{j\pi\Omega_l},\ldots,e^{j\pi(N-1)\Omega_l}\right]^{\rm T}\nonumber \\
&\triangleq \boldsymbol{\beta}(N,\Omega_l)\label{Eq.simplified steering vector}.
\end{align}
In fact, $\boldsymbol{\beta}(N,\Omega_l)$ is the far-field channel steering vector independent of $r_l$. 

If $r_l\le Z$, the approximation in \eqref{Eq.simplified steering vector} is not accurate enough. Alternatively, we approximate $r_l^{(n)}$ as  $r_l^{(n)} \approx r_l - \Omega_l\delta_n\lambda + \frac{\delta_n^2\lambda^2(1-\Omega_l^2)}{2r_l}$ according to $\sqrt{1+\epsilon}\approx1+\frac{1}{2}\epsilon - \frac{1}{8}\epsilon^2$, which is verified to be accurate if $r_l^{(n)}\ge 0.5\sqrt{D^3/\lambda}$~\cite{Tcom22CMH,APM17SKT}. Note that the ratio of $Z$ to $0.5\sqrt{D^3/\lambda}$ is $\sqrt{8N}$, which indicates that the latter is much smaller than the former, especially for XL-MIMO systems. For example, if $N = 512$, $0.5\sqrt{D^3/\lambda}$ is only $1/64$ of $Z$, which is nearly negligible. Therefore, in this work, we focus on the near field with $r_l^{(n)}\ge 0.5\sqrt{D^3/\lambda}$. Then $\boldsymbol{\alpha}(N,\Omega_l,r_l)$ in \eqref{Eq.steering vector} can be approximated as
\begin{align}
	\boldsymbol{\alpha}(\!N,\!\Omega_l,\!r_l) &\!\approx\! \frac{1}{\sqrt{N}}\!\big[e^{j2\pi(\Omega_l\delta_1\!-\!\rho_l\delta_1^2)}\!,\!\ldots\!,\!e^{j2\pi(\Omega_l\delta_N\!-\!\rho_l\delta_N^2)}\big]^{\rm T}
\end{align}	
where $\rho_l \triangleq \frac{\lambda(1-\Omega_l^2)}{2r_l}$. Define $b_l \triangleq \Omega_l + \rho_l(N+1)/2$ and $k_l \triangleq -\rho_l/2$. Multiplying a constant factor $e^{-j2\pi(\Omega_l(\delta_1-1/2) - \rho_l (N+1)^2/16)} $ to all the entries of $\boldsymbol{\alpha}(N,\Omega_l,r_l)$ in \eqref{FFSteeringVector}, we have
\begin{align}\label{chirpchannel}
	&~~~~e^{-j2\pi(\Omega_l(\delta_1-1/2) - \rho_l(N+1)^2/16)} \boldsymbol{\alpha}(N,\Omega_l,r_l) \nonumber \\
	&\approx \frac{1}{\sqrt{N}}\!\left[e^{j\pi(k_l+b_l)}\!,\!\ldots\!,e^{j\pi(k_ln^2+b_ln)},\ldots, \!e^{j\pi(k_lN^2+b_lN)}\right]^{\rm T}\nonumber \\
	&\triangleq \boldsymbol{\gamma}(N,\Omega_l,r_l).
\end{align}
Note that $\boldsymbol{\beta}(N,\Omega_l)$ is a special case of $\boldsymbol{\gamma}(N,\Omega_l,r_l)$ with~$\rho_l = 0$. Therefore, $\boldsymbol{\gamma}(N,\Omega_l,r_l)$ can be used  to approximate both the far-field and the near-field channel steering vectors. 


\section{Two-Stage Hybrid-Field Beam Training}\label{SecTSHFBT}
In this section, we will use the far-field channel steering vectors of a subarray to approximate its near-field channel steering vectors. Then we propose a THBT scheme for both the near field and far field.
\subsection{Hybrid-Field Beam Sweeping}\label{HFBSs}
To estimate the multipath channel, codebook-based beam training is widely adopted \cite{TWC20QCH}. Since the user is either in the far field or near field, a hybrid-field codebook considering both the  near-field and far-field effects will be developed in the following. First, we establish a far-field codebook $\boldsymbol{C}_{\rm f}\in\mathbb{C}^{N\times Q}$ based on the far-field channel steering vectors~\cite{TWC20QCH}, where $Q$ denotes the quantized number of the angle and the $q$th column of $\boldsymbol{C}_{\rm f}$ is denoted as
\begin{equation}\label{FarFieldCodebook}
	[\boldsymbol{C}_{\rm f}]_{:,q} \triangleq \boldsymbol{\beta}\left(N,\frac{2q-1-Q}{Q}\right)
\end{equation}
for $q=1,2,\ldots,Q$. Then we establish a near-field codebook $\boldsymbol{C}_{\rm n}\in\mathbb{C}^{N\times (QS)}$. Since the near-field channel is relevant to both the distance and the angle, we quantize the distance and the angle by $S$ samples and $Q$ samples, respectively. The $q$th angle sample is $\Theta_q = (2q-1-Q)/Q$ for $q=1,2,\ldots,Q$. According to the polar-domain sparsity in~\cite{Tcom22CMH}, the $s$th distance sample can be expressed as
\begin{equation}\label{Eq.quantizeddistance}
	d_{q,s} = \frac{N^2\lambda(1-\Theta_q^2)}{8\beta s}
\end{equation}
for $s=1,2,\ldots,S$,  where $\beta$ is a factor to adjust the coherence of the adjacent codewords. By setting ${N^2\lambda}/{(8\beta S)} = 0.5\sqrt{D^3/\lambda}$, we have $\beta = \sqrt{\frac{N}{2S^2}}$. Then \eqref{Eq.quantizeddistance} can be rewritten as
	\begin{align}
		d_{q,s} = \frac{N^{3/2}\lambda S(1-\Theta_q^2)}{4\sqrt{2}s}.
	\end{align} 
Let $\boldsymbol{C}_{\rm n} \triangleq \{ \boldsymbol{C}_{\rm 1},\boldsymbol{C}_{\rm 2},\ldots,\boldsymbol{C}_{Q} \}$, where the $s$th column of $[\boldsymbol{C}_{q}]$ is $[\boldsymbol{C}_{q}]_{:,s} = \boldsymbol{\alpha}(N,\Theta_q,d_{q,s})$. Accordingly, the hybrid-field codebook is defined as
\begin{align}\label{HybridFieldCodebook}
	\boldsymbol{C}_{\rm h} \triangleq \{\boldsymbol{C}_{\rm n},\boldsymbol{C}_{\rm f}\}\in\mathbb{C}^{N\times (QS+Q)}.
\end{align}
Based on \eqref{system model}, we have
\begin{equation}\label{Eq.beamtraining}
	y_{p} = [\boldsymbol{C}_{\rm h}]_{:,p}^{\rm H}\boldsymbol{h}x_{p} +[\boldsymbol{C}_{\rm h}]_{:,p}^{\rm H}\boldsymbol{\eta}
\end{equation}
for $p = 1,2,\ldots,QS+Q$. The beam training aims at finding the codeword in $\boldsymbol{C}_{\rm h}$ best fitting for the multipath channel, which can be expressed as
\begin{equation}\label{Eq.beamtrainingobjective}
	\widetilde{p} = \arg\max_{p=1,2,\ldots,QS+Q} \big|[\boldsymbol{C}_{\rm h}]_{:,p}^{\rm H}\boldsymbol{h}\big|.
\end{equation}
To solve \eqref{Eq.beamtrainingobjective}, we need to test all the codewords in $\boldsymbol{C}_{\rm h}$ one by one,  which is called the hybrid-field beam sweeping (HFBS). To perform the HFBS, $(QS+Q)$ times of beam training are needed. Comparing \eqref{Eq.simplified steering vector} with \eqref{Eq.steering vector}, larger overhead is needed by the near-field beam training than by the far-field beam training because the former needs $S$ times beam training for different distances even for the same angle while the latter needs only one time of beam training for the same angle. Therefore, it would be interesting to consider how to use the far-field beam training for the near-field channel, which will be discussed subsequently.
\subsection{Subarray Approximation}\label{SubarrayAppro}
We define an analog combiner as
\begin{equation}\label{AnalogCombinerBlockdiag}
	\boldsymbol{W} \triangleq {\rm blkdiag}\{\boldsymbol{w}_1,\boldsymbol{w}_2,\ldots,\boldsymbol{w}_{N_{\rm RF}}\}
\end{equation}
and a digital combiner as  $\boldsymbol{v} \in \mathbb{C}^{1\times N_{\rm RF}}$, where $\boldsymbol{w}_{t}\in\mathbb{C}^{1\times M}$  for $t = 1,2,\ldots,N_{\rm RF}$. Given $\boldsymbol{h}$, the optimal hybrid combiner to achieve the maximum received power can be designed via solving the problem
\begin{align}\label{Eq.beamtrainingProblem2}
\underset{\boldsymbol{W},\boldsymbol{v}}{\max}\ \ &\big| \boldsymbol{v}\boldsymbol{W}\boldsymbol{h}\big| \nonumber \\
\mathrm{s.t.\ }~~&\|\boldsymbol{v}\boldsymbol{W}\|_2=1, \big|[\boldsymbol{w}_t]_m\big|=1 
\end{align}
for $m=1,2,\ldots,M$ and $t=1,2,\ldots, N_{\rm RF}$. Due to the much smaller path gain of the NLoS paths than that of the LoS path especially in millimeter wave or terahertz band, we omit the NLoS paths. Then \eqref{Eq.beamtrainingProblem2} can be rewritten as
\begin{subequations}\label{Eq.beamtrainingProblem3} \normalsize
\begin{align}
\underset{\boldsymbol{W},\boldsymbol{v}}{\max}\ &\big| \boldsymbol{v}\boldsymbol{W}\boldsymbol{\alpha}(N,\Omega,r)\big| \label{Eq.beamtrainingObjective3} \\
\mathrm{s.t.}~~&\|\boldsymbol{v}\boldsymbol{W}\|_2=1,~\big|[\boldsymbol{w}_t]_m\big|=1 \label{envelop constrain3}
\end{align}
\end{subequations}
for $m=1,2,\ldots,M$ and $t=1,2,\ldots, N_{\rm RF}$. In \eqref{Eq.beamtrainingObjective3}, the subscript ``$l$" is omitted for simplicity, which goes the same for the rest of this paper.

According to the Cauchy-Schwartz inequality, we have
\begin{equation}\label{UpperBoundObjective}
	\big| \boldsymbol{v}\boldsymbol{W}\boldsymbol{\alpha}(N,\Omega,r)\big|\leq \|\boldsymbol{v}\|_2~ \|\boldsymbol{W}\boldsymbol{\alpha}(N,\Omega,r)\|_2.
\end{equation}
The equality in \eqref{UpperBoundObjective} holds if $\boldsymbol{v}^{\rm H}=\mu\boldsymbol{W}\boldsymbol{\alpha}(N,\Omega,r)$, where $\mu$ is a scaling factor. Since $\boldsymbol{W}$ and $\boldsymbol{v}$ are independent of each other, we can achieve the equality of \eqref{UpperBoundObjective}. Consequently, we may first determine $\boldsymbol{W}$ and then design $\boldsymbol{v}$ accordingly.

The design of $\boldsymbol{W}$ can be formulated as
\begin{align}\label{Eq.beamtrainingProblem4}
\underset{\boldsymbol{W}}{\max}\ &\big\|\boldsymbol{W}\boldsymbol{\alpha}(N,\Omega,r)\big\|_2 \nonumber\\
\mathrm{s.t.\ }~&\big|[\boldsymbol{w}_t]_m\big|=1 
\end{align}
for $m=1,2,\ldots,M$ and $t=1,2,\ldots, N_{\rm RF}$. The entries of $\boldsymbol{W}$ are mutually independent, implying that the maximization of $\|\boldsymbol{W}\boldsymbol{\alpha}(N,\Omega,r)\|_2$  is essentially the maximization of the absolute value of each entry of $\boldsymbol{W}\boldsymbol{\alpha}(N,\Omega,r)$. Therefore, the problem in \eqref{Eq.beamtrainingProblem4} is divided into $N_{\rm RF}$ independent subproblems and the $t$th subproblem for $t=1,2,\ldots,N_{\rm RF}$ can be expressed as
\begin{align}\label{Eq.beamtrainingProblem5}
\underset{\boldsymbol{w}_t}{\max}\ &\big |\boldsymbol{w}_t\boldsymbol{G}_t\boldsymbol{\alpha}(N,\Omega,r)\big | \nonumber\\
\mathrm{s.t.\ }~&\big|[\boldsymbol{w}_t]_m\big|=1
\end{align}
for $m=1,2,\ldots,M$, where we define  $\boldsymbol{G}_t \triangleq [\boldsymbol{0}_{M\times(t-1)M},\boldsymbol{I}_{M},\boldsymbol{0}_{M\times(N_{\rm RF}-t)M}]$. The optimal solution of \eqref{Eq.beamtrainingProblem5} is
\begin{equation}\label{Eq.OptimalSolution}
\overline{\boldsymbol{w}}_{t} =\sqrt{N} \big(\boldsymbol{G}_t\boldsymbol{\alpha}(N,\Omega,r)\big)^{\rm H}
\end{equation}
for $t=1,2,\ldots,N_{\rm RF}$.

Note that $\boldsymbol{G}_t\boldsymbol{\alpha}(N,\Omega,r)$ is exactly the channel steering vector between the user and the $t$th subarray. From \eqref{Rayleigh distance}, the entire array at the BS achieves larger $Z$ than each subarray. The Rayleigh distance decreases quadratically with the reduction of the antenna number. In fact, the Rayleigh distance of a subarray is only $1/N_{\rm RF}^2$ of the Rayleigh distance of the entire array at the BS, which implies that a user in the near field of the entire array may be in the far field of a subarray. For example, if $N=256$, $\lambda = 0.003~{\rm m}$, and $N_{\rm RF} = 4$, the Rayleigh distance of the entire array is 98.3m while the Rayleigh distance of a subarray is only 6.1m. In other words, the near-field effect of subarrays is much weaker than that of the entire array. Based on the above discussion, we try to use the far-field channel steering vectors to approximate the near-field steering vectors for subarrays. To evaluate the deviation of the approximation, we have the following lemmas proved in Appendices A and B.
	
\noindent\textbf{Lemma 1:} The maximum beam gain loss of approximating near-field channel steering vector with far-field channel steering vector for a  subarray can be approximated as
\begin{align}
	\Gamma_{\rm max} = \max\{{1-{N_{\rm RF}}/{\sqrt[4]{{2N}}}},0\}.
\end{align}
\noindent\textbf{Lemma 2:} Denote the sine result of the angle that points from the center of the $t$th subarray to the user as $\Psi_t$ and denote the beam center of $\boldsymbol{G}_t\boldsymbol{\alpha}(N,\Omega,r)$ as $B_t$. Then, we have $B_t\approx\Psi_t$.

According to \textbf{Lemma 1}, the beam gain loss of approximating the near-field channel steering vector with the far-field channel steering vector is limited. For example, if $N=256$ and $N_{\rm RF} = 4$, the maximum beam gain loss is only $16\%$. From \textbf{Lemma 2}, considerable beamforming gain can be obtained if the $t$th subarray receives the signal from channel $\boldsymbol{G}_t\boldsymbol{\alpha}(N,\Omega,r)$ with $\boldsymbol{\beta}(M,\Psi_t)$.

Based on \textbf{Lemma 1} and \textbf{Lemma 2}, for each subarray, we use far-field channel steering vectors,
\begin{equation}\label{Eq.SubOptimalSolution}
 \widehat{\boldsymbol{w}}_t= \sqrt{M}\boldsymbol{\beta}(M,\Psi_t)^{\rm H}
\end{equation}
for $t=1,2,\ldots,N_{\rm RF}$ to approximate the near-field channel steering vectors in \eqref{Eq.OptimalSolution}, which is termed the subarray approximation. Therefore, we can use \eqref{Eq.SubOptimalSolution} for both the far-field and near-field channels. Note that the subarray approximation has also been studied in the existing works, such as \cite{Arxiv21_CMY}. The existing works aim to design the wideband beamforming based on the ideal CSI. However, our work performs the effective beam training to obtain the CSI.

By substituting \eqref{Eq.SubOptimalSolution} into \eqref{AnalogCombinerBlockdiag}, we express the designed analog combiner for \eqref{Eq.beamtrainingProblem3} as
\begin{equation}\label{Analogbeamforming}
\widehat{\boldsymbol{W}} = {\rm blkdiag}\{\widehat{\boldsymbol{w}}_1,\widehat{\boldsymbol{w}}_2,\ldots,\widehat{\boldsymbol{w}}_{N_{\rm RF}}\}.
\end{equation}
Since $\widehat{\boldsymbol{W}}\widehat{\boldsymbol{W}}^{\rm H} = M\boldsymbol{I}_{N_{\rm RF}}$, we have $\|\boldsymbol{v}\widehat{\boldsymbol{W}}\|_2=\sqrt{M}\|\boldsymbol{v}\|_2$. Then the design of $\boldsymbol{v}$ according to \eqref{Eq.beamtrainingProblem3} can be expressed as
\begin{subequations}\label{Eq.beamtrainingProblem6} \normalsize
\begin{align}
\underset{\boldsymbol{v}}{\max}\ &\boldsymbol{v}\widehat{\boldsymbol{W}}\boldsymbol{\alpha}(N,\Omega,r) \label{Eq.beamtrainingObjective6} \\
\mathrm{s.t.}~~&\|\boldsymbol{v}\|_2=1/\sqrt{M}. \label{envelop constrain6}
\end{align}
\end{subequations}
Note that \eqref{Eq.beamtrainingObjective3} can be rewritten as \eqref{Eq.beamtrainingObjective6} because we can always adjust the phase of $\boldsymbol{v}$ so that $\boldsymbol{v}\widehat{\boldsymbol{W}}\boldsymbol{\alpha}(N,\Omega,r)$ is real positive and the maximum of $|\boldsymbol{v}\widehat{\boldsymbol{W}}\boldsymbol{\alpha}(N,\Omega,r)|$ is still achieved. The optimal $\boldsymbol{v}$ for \eqref{Eq.beamtrainingProblem6} is
\begin{equation}\label{Eq.OptimalfBB}
\widehat{\boldsymbol{v}} = \frac{\left(\widehat{\boldsymbol{W}}\boldsymbol{\alpha}(N,\Omega,r)\right)^{\rm H}}{\sqrt{M}\big\|\widehat{\boldsymbol{W}}\boldsymbol{\alpha}(N,\Omega,r)\big\|_2}.
\end{equation}

Note that  \eqref{Analogbeamforming}  and \eqref{Eq.OptimalfBB} are designed based on the continuous space. To facilitate the implementation of beam training, hybrid combiners for the quantized space need to be designed. For each subarray, the commonly-used DFT codebook for beam training is
\begin{equation}\label{Eq.SetofSteeringVector}
	\boldsymbol{\Phi}= \{\sqrt{M}\boldsymbol{\beta}(M,\Phi_1),\sqrt{M}\boldsymbol{\beta}(M,\Phi_2),\ldots,\sqrt{M}\boldsymbol{\beta}(M,\Phi_M)\},
\end{equation}
where
\begin{equation}
	\Phi_m = (2m-1-M)/M
\end{equation}
for $m=1,2,\ldots,M$. In fact, the DFT codebook equally samples the full space $[-1,1]$ by $M$ angles, where the $m$th angle is $\Phi_m$. For the $t$th subarray, the index of the codeword in $\boldsymbol{\Phi}$ best fitting for the channel is
\begin{equation}\label{Eq.Quantizedft2}
	\widetilde{m}_t = \arg\min_{m = 1,2,\ldots,M}|\Phi_m-\Psi_t|.
\end{equation}
Therefore, the designed analog combiner and digital combiner according to \eqref{Analogbeamforming} and \eqref{Eq.OptimalfBB}, respectively, are
\begin{align}
	&\widetilde{\boldsymbol{W}} = {\rm blkdiag}\{\widetilde{\boldsymbol{w}}_1,\widetilde{\boldsymbol{w}}_2,\ldots,\widetilde{\boldsymbol{w}}_{N_{\rm RF}}\}, \label{Eq.QuantizedFrf}\\
	&\widetilde{\boldsymbol{v}} = \frac{\left(\widetilde{\boldsymbol{W}}\boldsymbol{\alpha}(N,\Omega,r)\right)^{\rm H}}{\sqrt{M}\big\|\widetilde{\boldsymbol{W}}\boldsymbol{\alpha}(N,\Omega,r)\big\|_2},\label{Eq.QuantizedFbb}
\end{align}
where
\begin{equation}\label{Eq.Quantizedft}
 \widetilde{\boldsymbol{w}}_t= [\boldsymbol{\Phi}]_{:,\widetilde{m}_t}^{\rm H}
\end{equation}
for $t=1,2,,\ldots,N_{\rm RF}$ based on \eqref{Eq.SubOptimalSolution}.

\begin{figure}[!t]
	\begin{center}
		\includegraphics[width=75mm]{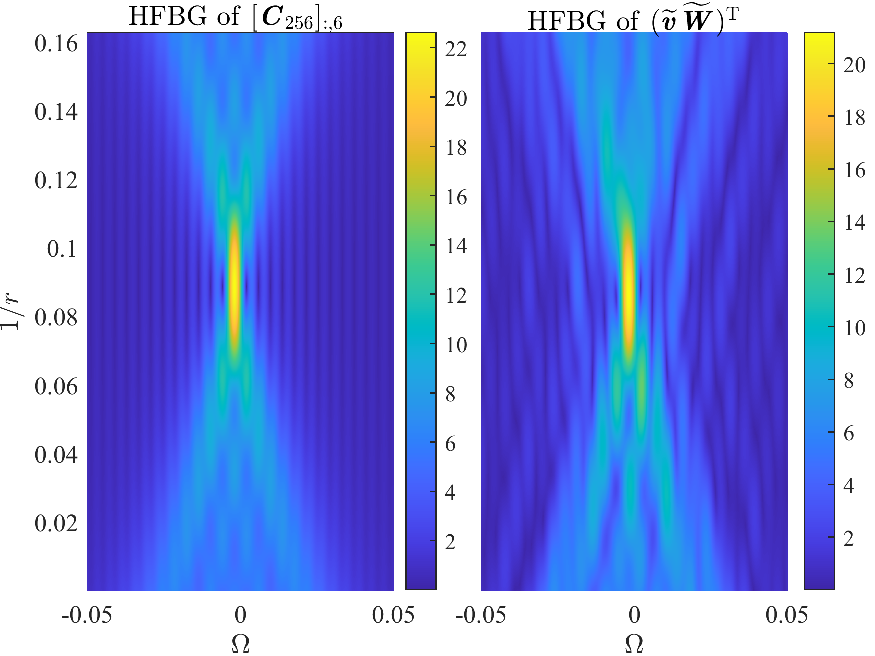}
	\end{center}
	\caption{Illustration of the subarray approximation.}	\label{SubarrayApproximation}
\end{figure}

Fig.~\ref{SubarrayApproximation} illustrates the performance of the subarray approximation, where $N = 512$, $N_{\rm RF} = 4$, $Q = 512$ and $S = 11$. We take the subarray approximation of $[\boldsymbol{C}_{256}]_{:,6} = \boldsymbol{\alpha}(N,\Theta_{256},d_{256,6})$ as an example. Based on \eqref{Eq.Quantizedft2}, we can obtain $\widetilde{m}_1 = 63$, $\widetilde{m}_2 = 64$, $\widetilde{m}_3 = 65$ and $\widetilde{m}_4 = 66$. The analog combiner and the digital combiner can be designed based on \eqref{Eq.QuantizedFrf} and \eqref{Eq.QuantizedFbb}, respectively. Then the combined channel steering vector is $(\widetilde{\boldsymbol{v}}\widetilde{\boldsymbol{W}})^{\rm T}$. We define the hybrid-field beam gain (HFBG) of a channel steering vector $\boldsymbol{u}$ as 
\begin{align}
B(\boldsymbol{u},\Omega,r) = |\boldsymbol{\alpha}(N,\Omega,r)^{\rm H}\boldsymbol{u}|.
\end{align}
In Fig.~\ref{SubarrayApproximation}, we illustrate the HFBGs of $[\boldsymbol{C}_{256}]_{:,6}$ and $(\widetilde{\boldsymbol{v}}\widetilde{\boldsymbol{W}})^{\rm T}$. From the figure, the HFBG of $(\widetilde{\boldsymbol{v}}\widetilde{\boldsymbol{W}})^{\rm T}$ is similar to that of $[\boldsymbol{C}_{256}]_{:,6}$, which verifies the effectiveness of the subarray approximation.

\subsection{Beam Training}
Now we propose a two-stage hybrid-field beam training scheme.

Note that for the partially-connected structure, each subarray is exclusively connected to an RF chain, which indicates that each RF chain can support an independent beam training based on a subarray. In addition, the near-field effect is substantially weakened once the ULA is divided into $N_{\rm RF}$ subarrays. Therefore, for each angle, we only need one time of beam training no matter whether the user is in the near field or far field.
 
Based on the above discussions, in the first stage, each subarray  independently performs beam training with codewords in $\boldsymbol{\Phi}$. When performing the $m$th beam training,  the analog combiner according to \eqref{Eq.SetofSteeringVector} can be expressed as
\begin{equation}\label{Eq.FrfDurBeamTraining}
	\overline{\boldsymbol{W}}_{m} = {\rm blkdiag}\{\sqrt{M}\boldsymbol{\beta}(M,\Phi_m)^{\rm H},\ldots,\sqrt{M}\boldsymbol{\beta}(M,\Phi_m)^{\rm H}\}.
\end{equation}
for $m = 1,2,\ldots,M$. Based on \eqref{system model}, the output signal of the $k$th combiner is
\begin{equation}\label{Eq.ReceivedSignalBDC}
	\boldsymbol{z}_m = \overline{\boldsymbol{W}}_{m}\boldsymbol{h}x_{m} + \overline{\boldsymbol{W}}_{m}\boldsymbol{\eta}
\end{equation}
for $m=1,2,\ldots,M$.

Note that according to the analysis in Section \ref{SubarrayAppro}, the received signals of subarrays can be combined so that the power of combined signals can approximate the power of signals received by the entire array. In addition, for any codeword in $\boldsymbol{C}_{\rm h}$, the analog combiners of subarrays come from $\boldsymbol{\Phi}$. Moreover, all subarrays have performed beam training with $\boldsymbol{\Phi}$ in the first stage. Therefore, by reusing the received signals in the first stage and designing the digital combiners, the  power of the combined signals can approximate the power of signals received by codewords in $\boldsymbol{C}_{\rm h}$.

Based on the above discussions, in the second stage, by utilizing $\{\boldsymbol{z}_1,\boldsymbol{z}_2,\ldots,\boldsymbol{z}_M\}$, we design the digital combiner, $\boldsymbol{v}_p$, for $p = 1,2,\ldots,QS+Q$,  to test all the codewords in the hybrid-field codebook $\boldsymbol{C}_{\rm h}$. To be detailed, for each codeword $[\boldsymbol{C}_{\rm h}]_{:,p}$, we can design a dedicated digital combiner $\boldsymbol{v}_p$ to combine the output of the analog combiner from the first stage.  From \eqref{HybridFieldCodebook}, we~have
\begin{equation}\label{pthcodeword}
[\boldsymbol{C}_{\rm h}]_{:,p} = \left\{ \begin{array}{ll}
\boldsymbol{\alpha}\left(N,\Theta_{\bar{q}},d_{\bar{q},\bar{s}}\right), & p \leq QS,\\
\boldsymbol{\beta}\left(N,\frac{2(p-QS)-1-Q}{Q}\right), & p > QS,
\end{array} \right.
\end{equation}
where
\begin{equation}\label{barn}
\bar{q} = \big\lceil p/S\big\rceil~\mbox{and}~\bar{s} = p - (\bar{q}-1)S.
\end{equation}
\begin{figure}[!t]
	\begin{center}
		\includegraphics[width=88mm]{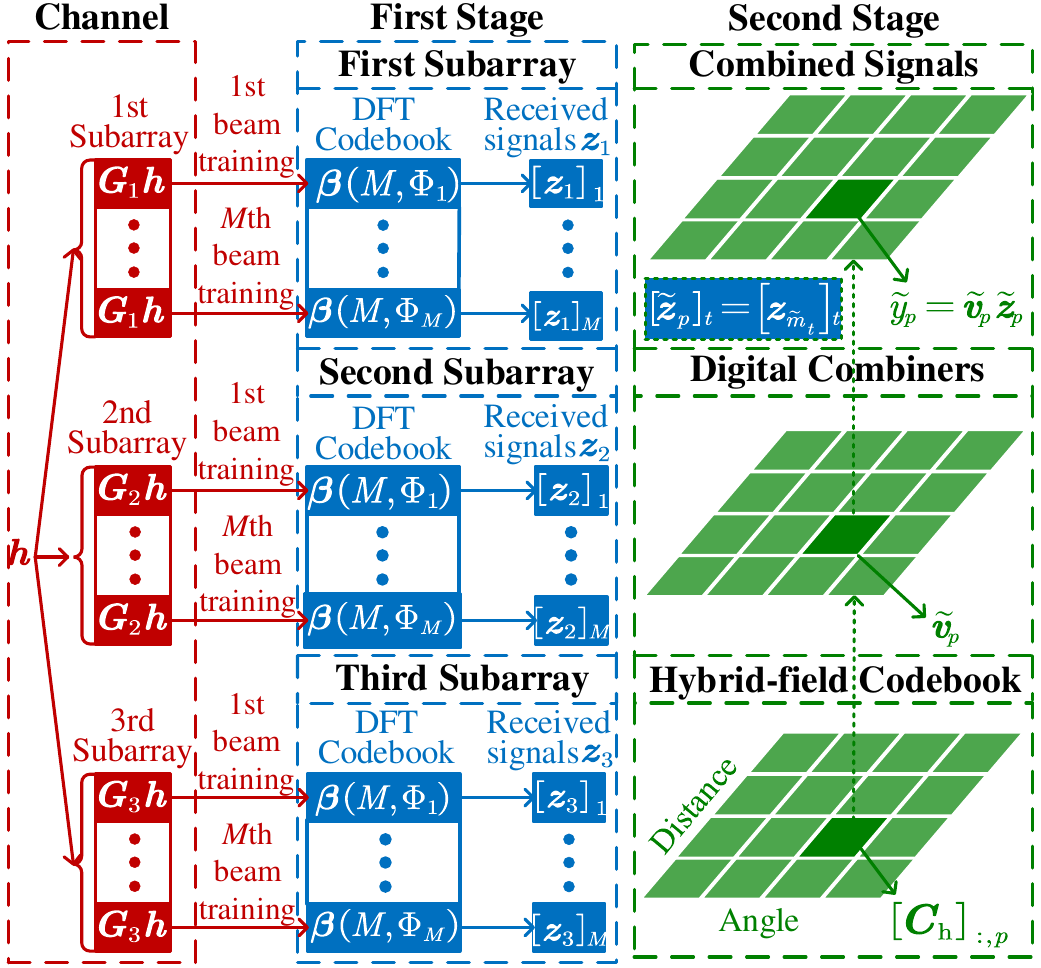}
	\end{center}
\caption{Illustration of the procedure for the THBT scheme.}\label{Signaling}
		\vspace{-0.5cm}
\end{figure}
Replacing $\boldsymbol{\alpha}(N,\Omega,r)$ in \eqref{Eq.beamtrainingProblem3} by $[\boldsymbol{C}_{\rm h}]_{:,p}$, we can obtain $\widetilde{\boldsymbol{w}}_t^{(p)}$ via \eqref{Eq.Quantizedft}. Similar to \eqref{Eq.QuantizedFrf} and \eqref{Eq.QuantizedFbb}, we design the analog combiner and the digital combiner, respectively, as
\begin{align}
&\boldsymbol{F}_p = {\rm blkdiag}\{\widetilde{\boldsymbol{w}}_1^{(p)},\widetilde{\boldsymbol{w}}_2^{(p)},\ldots,\widetilde{\boldsymbol{w}}_{N_{\rm RF}}^{(p)}\}, \label{Eq.QuantizedFrfCh}\\
&\widetilde{\boldsymbol{v}}_{p} = \frac{\left({\boldsymbol{F}}_p[\boldsymbol{C}_{\rm h}]_{:,p}\right)^{\rm H}}{\sqrt{M}\big\|{\boldsymbol{F}}_p[\boldsymbol{C}_{\rm h}]_{:,p}\big\|_2}.\label{Eq.QuantizedFbbCh}
\end{align}
If we use $\boldsymbol{F}_p$ and $\widetilde{\boldsymbol{v}}_{p}$ for combining, we have
\begin{equation}\label{Eq.ReceivedSignalBDC2}
	\widetilde{\boldsymbol{z}}_{p} = \boldsymbol{F}_{p}\boldsymbol{h}x_p + \boldsymbol{F}_{p}\boldsymbol{\eta},
\end{equation}
and
\begin{equation}\label{Eq.ReceivedSignalADC}
\widetilde{y}_{p} = \widetilde{\boldsymbol{v}}_{p} \widetilde{\boldsymbol{z}}_{p},~p = 1,2,\ldots,QS+Q.
\end{equation}
In fact, $\widetilde{\boldsymbol{z}}_{p}$ can not be obtained directly because we do not actually perform beam training with  $\boldsymbol{F}_{p}$. However, each entry of $\widetilde{\boldsymbol{z}}_{p}$ can be obtained from the beam training in \eqref{Eq.ReceivedSignalBDC} because both ${\boldsymbol{F}}_{p}$ and $\overline{\boldsymbol{W}}_{m}$ are composed of channel steering vectors from the same set $\boldsymbol{\Phi}$ in \eqref{Eq.SetofSteeringVector}.  We can obtain $\widetilde{\boldsymbol{z}}_{p}$  by setting
\begin{equation}\label{zpt}
[\widetilde{\boldsymbol{z}}_{p}]_t = [\boldsymbol{z}_{\widetilde{m}_t}]_t,
\end{equation}
where $\widetilde{m}_t$ can be obtained from \eqref{Eq.Quantizedft2} during the design of $\boldsymbol{F}_p$ in \eqref{Eq.QuantizedFrfCh}. From $\{\widetilde{y}_{1},\widetilde{y}_{2},\ldots,\widetilde{y}_{QS+Q}\}$, we select the one with the largest power, which can be expressed as
\begin{equation}\label{Eq.beamtrainingresult}
  \widetilde{p} = \arg\max_{p = 1,2,\ldots,QS+Q} \big|\widetilde{y}_{p}\big|^2.
\end{equation}
\begin{algorithm}[!t]
	\caption{Two-Stage Hybrid-Field Beam Training (THBT) Scheme}
	\label{alg-TSBT}
	\begin{algorithmic}[1]
		\STATE \textbf{Input:} $N$, $N_{\rm RF}$, $M$, $Q$, $S$, $\lambda$.
		\STATE \textbf{First Stage:}
		\STATE Obtain $\boldsymbol{z}_m,~m=1,2,\ldots,M$ via \eqref{Eq.ReceivedSignalBDC}.
		\STATE \textbf{Second Stage:}
		\FOR{$p=1,2,\ldots, QS+Q$}
		\STATE Obtain $ [\boldsymbol{C}_{\rm h}]_{:,p}$ via \eqref{pthcodeword}.
		\STATE Obtain $\boldsymbol{F}_p$ and $\widetilde{\boldsymbol{v}}_{p}$ via \eqref{Eq.QuantizedFrfCh} and \eqref{Eq.QuantizedFbbCh}, respectively.
		\STATE Obtain $\widetilde{\boldsymbol{z}}_{p}$ via \eqref{zpt}.
		\STATE Obtain $\widetilde{{y}}_{p}$  via \eqref{Eq.ReceivedSignalADC}.
		\ENDFOR
		\STATE Obtain $\widetilde{p}$ via \eqref{Eq.beamtrainingresult}.
		\STATE Obtain $\widetilde{\Omega}$ and $\widetilde{r}$ via \eqref{roughestimation}.
		\STATE \textbf{Output:} $[\boldsymbol{C}_{\rm h}]_{:,\widetilde{p}}$, $\widetilde{\Omega}$ and $\widetilde{r}$.
	\end{algorithmic}
\end{algorithm}
From the hybrid-field codebook, $\boldsymbol{C}_{\rm h}$, we select the codeword $[\boldsymbol{C}_{\rm h}]_{:,\widetilde{p}}$ corresponding to the dedicated digital combiner $\widetilde{\boldsymbol{v}}_{\widetilde{p}} $ that can achieve the largest combining power $|\widetilde{y}_{\widetilde{p}}|^2$ as the result of the THBT.

Furthermore, if the user is in the near field, we can also roughly locate the user based on the beam training results. According to \eqref{barn}, the indices of the quantized angle and the quantized distance can be calculated as 
\begin{align}
\widetilde{q} = \big\lceil \widetilde{p}/S\big\rceil~\mbox{and}~\widetilde{s} = \widetilde{p} - (\widetilde{q}-1)S.
\end{align}
Then the angle and the distance of the user can be roughly estimated as 
\begin{align}\label{roughestimation}
	\widetilde{\Omega} = (2\widetilde{q}-1-Q)/Q~\mbox{and}~\widetilde{r} =  \frac{N^{3/2}\lambda S(1-\Theta_{\widetilde{q}}^2)}{4\sqrt{2}\widetilde{s}}.
\end{align}


Finally, The detailed steps of the proposed THBT scheme are summarized in \textbf{Algorithm~\ref{alg-TSBT}}.

Fig.~\ref{Signaling} illustrates the procedure for the THBT scheme, where $N_{\rm RF} = 3$. First, each subarray receives the uplink pilots independently based on the DFT codebook. Then, we design the digital combiners to combine the outputs of subarrays to test codewords in the predefined codebook. Finally, from the predefined hybrid-field codebook, we select the codeword that achieves the largest combining power as the result of the THBT.

Now we evaluate the overhead of the THBT scheme. The THBT scheme performs digital combining in the second stage, which does not need any pilots. Before digital combining, in the first stage, each subarray independently performs beam training based on the DFT codebook  and needs $M$ times of beam training. Therefore, the total overhead of the THBT scheme is $M$.

\section{Beam Refinement based on Phase Shifts of Subarrays}\label{SecBRPSS}
In this section, we focus on refining the quantized beam training results of the THBT. Based on the PSP and the time-frequency duality, the expressions of the subarray signals after analog combining are analytically derived and a BRPSS scheme with closed-form solutions is proposed for high-resolution channel parameter estimation.

After the THBT, the initial estimation of $k$ and $b$ in \eqref{chirpchannel} can be expressed as 
\begin{align}\label{wtkb}
	\widetilde{k} = -\frac{\lambda(1-\widetilde{\Omega}^2)}{4\widetilde{r}}~\mbox{and}~\widetilde{b} = \widetilde{\Omega} - \widetilde{k}(N+1).
\end{align}
Note that the resolutions of $\widetilde{k}$ and $\widetilde{b}$ are limited by the quantization of the hybrid codebook. To improve the estimation accuracy of $k$ and $b$, the user continues to transmit one uplink pilot and the BS receives that with the analog combiner
\begin{equation}
	\overline{\overline{\boldsymbol{W}}} = {\rm blkdiag}\{\overline{\overline{\boldsymbol{w}}}_1^{\rm H},\overline{\overline{\boldsymbol{w}}}_2^{\rm H},\ldots,\overline{\overline{\boldsymbol{w}}}_{N_{\rm RF}}^{\rm H}\},
\end{equation}
where 
\begin{align}
&[\overline{\overline{\boldsymbol{w}}}_t]_m = e^{j\pi(\widetilde{k}((t-1)M + m)^2 + \widetilde{b}((t-1)M + m))}
\end{align} 
for $m = 1,2,\ldots,M$ and $t = 1,2, \ldots,N_{\rm RF}$. Note that $\overline{\overline{\boldsymbol{W}}}$ is designed based on $\widetilde{k}$ and $\widetilde{b}$. Thus, the near-field effects can be alleviated and directional beams are formed to provide high beamforming gains.  Then the received signal of the $t$th subarray after analog combining can be expressed as
\begin{align}\label{SARS}
	\widehat{z}_t &\overset{\rm (a)}{\approx} \sum_{n = (t-1)M+1}^{tM}e^{-j\pi(\widetilde{k}n^2 + \widetilde{b}n)}[\boldsymbol{h}]_n \nonumber \\
	&\overset{\rm (b)}{\approx} g \sum_{n = (t-1)M+1}^{tM}e^{-j\pi(\widetilde{k}n^2 + \widetilde{b}n)}[\boldsymbol{\alpha}(N,\Omega,r)]_n                                                                                \nonumber \\
	&\overset{\rm (c)}{\approx}\widetilde{g} \sum_{n = (t-1)M+1}^{tM}e^{-j\pi(\widetilde{k}n^2 + \widetilde{b}n)} [\boldsymbol{\gamma}(N,\Omega,r)]_n                                              \nonumber \\
	&= \widetilde{g}\sum_{n = (t-1)M+1}^{tM}e^{j\pi(\Delta{k}n^2 + \Delta{b}n)}                                           \nonumber \\
	&= \widetilde{g}\sum_{m =1}^{M}e^{j\pi(\Delta{k}(m+(t-1)M)^2 + \Delta{b}(m+(t-1)M))}                                    \nonumber \\
	&= \widetilde{g}  e^{j\pi(\Delta k (t-1)^2M^2 + \Delta{b}(t-1)M)}        													\nonumber \\
	& ~~~~~~~~~~~~~~~~\cdot\sum_{m =1}^{M}e^{j\pi(\Delta{k}m^2 + \Delta b m + 2\Delta kmM(t-1))},
\end{align}
for $t = 1,2,\ldots,N_{\rm RF}$, where $\Delta{k} \triangleq k - \widetilde{k}$, $\Delta{b} \triangleq b - \widetilde{b}$ and $\widetilde{g} \triangleq ge^{j2\pi(\Omega(\delta_1-1/2) - \rho (N+1)^2/16)}$. In \eqref{SARS}, we omit the noise term and set $x_p = 1$ in $\rm (a)$, omit the effects of the NLoS paths in $\rm (b)$, and approximate $\boldsymbol{\alpha}(N,\Omega,r)$ with $\boldsymbol{\gamma}(N,\Omega,r)$ in $\rm (c)$ to simplify the analysis. The summation in \eqref{SARS} is tricky due to the quadratic phase term $\Delta k m^2$. To obtain a deeper insight of $\widehat{z}_t$, we first focus on $e^{j\pi\Delta k m^2}$. Define $\boldsymbol{s}\triangleq [e^{j\pi\Delta k},e^{j\pi4\Delta k},\ldots,e^{j\pi M^2\Delta k}]^{\rm T}$. According to the PSP~\cite{AMMSE,ASF,AMFI}, the discrete-time Fourier transform of  $\boldsymbol{s}$ is expressed~as 
\begin{align}
	\mathcal{F}(\omega) &= \sum_{m=1}^M [\boldsymbol{s}]_me^{-j\pi m \omega} \nonumber \\
						&\approx\left\{ \begin{array}{ll}
							\!e^{-j\pi(\frac{\omega^2}{4\Delta k} + \frac{1}{4})}\sqrt{\frac{1}{-\Delta k}},\!&\!2\Delta kM\!\leq\!\omega\!\leq\!2\Delta k,\\
							0, &\!\mbox{others},
						\end{array} \right.
\end{align}
where we assume $\Delta k < 0$ without loss of generality. Then, according to the time-frequency duality, we have 
\begin{align}\label{sm2}
	[\boldsymbol{s}]_m &= e^{j\pi\Delta k m^2}            \nonumber \\
	& \approx \frac{1}{2}\int_{-1}^1 \mathcal{F}(\omega) e^{j\pi m \omega} {\rm d}\omega \nonumber \\
	& = \frac{1}{2}\int_{2\Delta kM}^{2\Delta k} e^{-j\pi(\frac{\omega^2}{4\Delta k} + \frac{1}{4})}\sqrt{\frac{1}{-\Delta k}} e^{j\pi m \omega} {\rm d}\omega.
\end{align}
Substituting \eqref{sm2} into \eqref{SARS}, we have \eqref{approxspe}, which is shown at the top of the next page. 
\begin{figure*}[htbp]
	\begin{align}\label{approxspe}
		\widehat{z}_t &\approx \widetilde{g}  e^{j\pi(\Delta k (t-1)^2M^2 + \Delta{b}(t-1)M)} \cdot \sum_{m =1}^{M}\left(\frac{1}{2}\int_{2\Delta kM}^{2\Delta k} e^{-j\pi(\frac{\omega^2}{4\Delta k} + \frac{1}{4})}\sqrt{\frac{1}{-\Delta k}} e^{j\pi m \omega} {\rm d}\omega\right)  e^{j\pi(\Delta b m + 2\Delta kmM(t-1))} \nonumber \\
		&= \widetilde{g}  e^{j\pi(\Delta k (t-1)^2M^2 + \Delta{b}(t-1)M)}\frac{1}{2}\int_{2\Delta kM}^{2\Delta k} e^{-j\pi(\frac{\omega^2}{4\Delta k} + \frac{1}{4})}\sqrt{\frac{1}{-\Delta k}} \left(\sum_{m =1}^{M}  e^{j\pi(\Delta b m + 2\Delta kmM(t-1))}	e^{j\pi m w} \right) {\rm d}\omega \nonumber \\
		&=  \widetilde{g}  e^{j\pi(\Delta k (t-1)^2M^2 + \Delta{b}(t-1)M)}\frac{1}{2}\sqrt{\frac{1}{-\Delta k}} \int_{2\Delta kM}^{2\Delta k} e^{-j\pi(\frac{\omega^2}{4\Delta k} + \frac{1}{4})}\left( e^{j(M+1)(\pi\phi_t+\pi \omega)/2}\frac{\sin(M(\pi\phi_t 
			+ \pi \omega)/2)}{\sin((\pi\phi_t 
			+ \pi \omega)/2)}\right) {\rm d}\omega \nonumber \\
		&= \overline{g} e^{j\pi(\Delta \widetilde{k} (t-1)^2 + \Delta\widetilde{b}(t-1))} \int_{2\Delta kM}^{2\Delta k} \underset{\mathcal{P}(\omega)}{\underbrace{e^{j\pi\left(\frac{(M+1)\omega}{2} - \frac{\omega^2}{4\Delta k} \right)}}}\underset{\mathcal{A}(\phi_t,\omega)}{\underbrace{\frac{\sin(M(\pi\phi_t 
					+ \pi \omega)/2)}{\sin((\pi\phi_t 
					+ \pi \omega)/2)}}} {\rm d}\omega\nonumber \\
		&= \overline{g} \underset{\mathcal{C}(t)}{\underbrace{e^{j\pi(\Delta \widetilde{k} (t-1)^2 + \Delta\widetilde{b}(t-1))}}} \underset{\mathcal{B}(\phi_t)}{\underbrace{\int_{2\Delta kM}^{2\Delta k} \mathcal{P}(\omega)\mathcal{A}(\phi_t,\omega){\rm d}\omega}}\nonumber \\
		&= \overline{g} \mathcal{C}(t)\mathcal{B}(\phi_t).
	\end{align}
	\hrulefill
	\vspace{-0.3cm}
\end{figure*}
In \eqref{approxspe}, we have 
\begin{align}\label{SARSVar}
	&\overline{g}  \triangleq \frac{1}{2\sqrt{-\Delta k}}\widetilde{g}e^{j\pi((M+1)\Delta b/2 -1/4)},\nonumber \\
	&\Delta \widetilde{k} \triangleq \Delta kM^2,\nonumber \\
	&\Delta \widetilde{b} \triangleq (\Delta{b} + \Delta k(M +1))M, \nonumber \\
	&\phi_t \triangleq \Delta{b} + 2\Delta kM(t-1).
\end{align}

\textbf{Remark 1:} In this part, we focus on providing high beamforming gains for all subarrays by analyzing the relations between the space quantization in \eqref{Eq.quantizeddistance} and the received signals of subarrays in \eqref{approxspe}. From \eqref{approxspe}, the power of $\widehat{z}_t$ is determined by the value of $|\mathcal{B}(\phi_t)|$. Moreover, $\mathcal{B}(\phi_t)$ is essentially the summation of $\mathcal{A}(\phi_t,\omega)$ with weighted factors $\mathcal{P}(\omega)$. To provide high beam gains for all subarrays, the maximum peak shift of $\mathcal{A}(\phi_t,\omega)$ should be less than the half-power width so that each term of $\mathcal{A}(\phi_t,(\omega))$ can provide high beam gain for summation and the maximum phase shift of $\mathcal{P}(\omega)$ should be less than $\pi/2$ so that $\mathcal{A}(\phi_t,\omega)$ for different $w$ will not cancel each other. Then we have

\begin{subequations}
\begin{align}
	&|\phi_t + \omega| \le \frac{1}{M},  \label{SAD11}\\
    &\left|\frac{(M+1)\omega}{2} - \frac{\omega^2}{4\Delta k}\right|\le 1/2 \label{SAD12},
\end{align}
\end{subequations}
for $\omega \in [2\Delta k M,2\Delta k]$. Substituting $\phi_t$ in \eqref{SARSVar} into \eqref{SAD11}, we have 
\begin{align}\label{SAD21}
|\Delta k| \le \frac{1}{M(N-1)} -	\frac{1}{Q(N-1)}.
\end{align}
From \eqref{SAD12}, we have 
\begin{align}\label{SAD22}
 	|\Delta k|\le \frac{1}{4M^2}.
\end{align}
Note that 
\begin{align}
\frac{1}{M(N-1)} -	\frac{1}{Q(N-1)} < \frac{1}{M(N-1)}\approx \frac{1}{N_{\rm RF}M^2} \le\frac{1}{4M^2}
\end{align}
for $N_{\rm RF} \ge4$, which is easy to be satisfied for XL-MIMO systems. Therefore, we omit the constraint in \eqref{SAD22} and focus on \eqref{SAD21} to simplify the expression. Note that 
\begin{align}\label{MaxDevik}
	|\Delta k| &\le \frac{1}{4}\left|\frac{\lambda(1-\Theta_q^2)}{2d_{q,s}}-\frac{\lambda(1-\Theta_q^2)}{2d_{q,s+1}}\right| \nonumber \\
	&= \frac{\sqrt{2}}{2N^{3/2} S}.
\end{align}
Combining \eqref{SAD21} and \eqref{MaxDevik}, we have 
\begin{align}\label{MinS}
	Q\ge M,~~\mbox{and}~~~S \ge \frac{\sqrt{2}(N-1)}{2N^{3/2}\left(\frac{1}{M}-\frac{1}{Q}\right)},
\end{align}
which provides guidance for the settings of $S$ and $Q$.

In \eqref{approxspe}, the expression of $\widehat{z}_t$ is analytically derived, where the phase of  $\widehat{z}_t$  shifts along with $t$. Specifically, the phase of $\widehat{z}_t$ is related to $\overline{g}$, $\mathcal{C}(t)$, and $\mathcal{B}(\phi_t)$, where $\overline{g}$ is a constant irrelevant to $t$. In addition, the phase of $\mathcal{B}(\phi_t)$ is also irrelevant to $t$ because $\mathcal{P}(\omega)$ is  irrelevant to $t$ and $\mathcal{A}(\phi_t,\omega)$ consistently takes on positive values under the constraint in \eqref{SAD11}. Therefore, the unwrapped phase of $\widehat{z}_t$ is a quadratic function of $t$ with $\Delta kM^2$ and $(\Delta{b} + \Delta k(M +1))M$ as its coefficients, which indicates that the parameters of the channels are estimated if we can obtain the coefficients of the quadratic function. 

Define $\varUpsilon_t \triangleq \angle \widehat{z}_t$, where $\angle(\cdot)$ denotes the phase of a complex value. The first-order difference of $\varUpsilon_t$ can be expressed~as
\begin{align}\label{FoD}
	\Delta \varUpsilon_t &\triangleq \varUpsilon_{t+1} - \varUpsilon_{t} =  \pi (\Delta \widetilde{k}(2t-1)  + \Delta \widetilde{b}),
\end{align}
for $t = 1,2,\ldots,N_{\rm RF}-1$. Then the second-order difference of $\varUpsilon_t$ can be expressed as
\begin{align}\label{SoD}
	\Delta^2{\varUpsilon}_t &\triangleq \Delta\varUpsilon_{t+1} - \Delta\varUpsilon_{t} =  2\pi\Delta \widetilde{k},
\end{align}
for $t = 1,2,\ldots,N_{\rm RF}-2$. Note that 
\begin{align}
	|\Delta^2{\varUpsilon}_t|= 2\pi|\Delta \widetilde{k}| \overset{\mbox{(a)}}{<} \frac{2\pi M^2}{M(N-1)}\! =\!  \frac{2\pi}{N_{\rm RF}}\cdot \frac{N}{N-1}\overset{\mbox{(b)}}{<}\pi
\end{align}
where $\mbox{(a)}$ can be obtained via \eqref{SAD21} and $\mbox{(b)}$ holds for $N_{\rm RF}> 2$. To avoid phase wrap, we can shift $\Delta^2{\varUpsilon_t}$ to $[-\pi,\pi]$ by
\begin{align}
	\overline{\varUpsilon}_t  \triangleq \mbox{mod}(\Delta^2{\varUpsilon}_t+\pi,2\pi) - \pi.
\end{align}
Then the estimation of $\Delta k$ can be expressed as 
\begin{align}\label{widehatk}
	\Delta \widehat{k}  = \frac{1}{N_{\rm RF}-2}\sum_{t = 1}^{N_{\rm RF}-2} \frac{\overline{\varUpsilon}_t}{2\pi M^2}.
\end{align}
\begin{algorithm}[!t]
	\caption{Beam Refinement based on Phase Shifts of Subarrays  (BRPSS) Scheme}
	\label{BRPSS}
	\begin{algorithmic}[1]
		\STATE \textbf{Input:} $N$, $N_{\rm RF}$, $M$, $\lambda$, $\widehat{z}_t$, $\widetilde{\Omega}$ and $\widetilde{r}$.
		\STATE Obtain $\widetilde{k}$ and $\widetilde{b}$ via \eqref{wtkb}.
		\STATE Obtain $\varUpsilon_t = \angle\widehat{z}_t$, for $t =1,2,\ldots,N_{\rm RF}$.
		\STATE Obtain $\Delta \varUpsilon_t$ and $\Delta^2{\varUpsilon}_t$ via \eqref{FoD} and \eqref{SoD}, respectively.
		\STATE Obtain $\Delta \widehat{k}$ and $\Delta \widehat{b}$ via  \eqref{widehatk} and \eqref{widehatb}, respectively.
		\STATE Obtain $\widehat{k}$ and $\widehat{b}$ via \eqref{Refinedkb}.
		\STATE Obtain $\widehat{\Omega}$ and $\widehat{r}$ via \eqref{RefinedOmegaR}.
		\STATE \textbf{Output:} $\widehat{k}$, $\widehat{b}$, $\widehat{\Omega}$ and $\widehat{r}$.
	\end{algorithmic}
\end{algorithm}

Now we turn to the estimation of $\Delta b$. Define 
\begin{align}
\widetilde{\varUpsilon}_t  \triangleq \Delta\varUpsilon_t - (2t-1)M^2\Delta \widehat{k}\pi - M(M+1)\Delta \widehat{k}\pi.
\end{align}
Then according to \eqref{FoD}, we have 
\begin{align}\label{wpt}
	\widetilde{\varUpsilon}_t = M\Delta b\pi + 2u\pi,~u\in\mathbb{N}.
\end{align}
Note that
\begin{align}
	|M\Delta b\pi|\le \frac{M\pi}{Q} + \frac{(N+1)\pi}{(N-1)} -	\frac{M(N+1)\pi}{Q(N-1)}\overset{\mbox{(a)}}{\approx}\pi.
\end{align}
where $\mbox{(a)}$ holds because $(N+1)/(N-1)\approx 1$ for XL-MIMO systems. Therefore, we can also shift the results in \eqref{wpt} to $[-\pi,\pi]$  to avoid phase wrap. Define  $\widehat{\varUpsilon}_t \triangleq \mbox{mod}(\widetilde{\varUpsilon}_t+\pi,2\pi) - \pi$. Then the estimation of $\Delta b$ can be expressed as 
\begin{align}\label{widehatb}
	\Delta \widehat{b}  = \frac{1}{N_{\rm RF}-1}\sum_{t = 1}^{N_{\rm RF}-1} \frac{\widehat{\varUpsilon}_t}{M\pi}.
\end{align}
Then the refined estimation of $k$ and $b$ can be expressed as 
\begin{align}\label{Refinedkb}
	\widehat{k} = \Delta \widehat{k} + \widetilde{k}~\mbox{and}~\widehat{b} = \Delta \widehat{b} + \widetilde{b}.
\end{align}
Accordingly, the refined estimation of $r$ and $\Omega$ can be expressed as 
\begin{align}\label{RefinedOmegaR}
	\widehat{\Omega} = \widehat{b} + \widehat{k}(N+1)~\mbox{and}~\widehat{r} = -\frac{\lambda(1-\widehat{\Omega}^2)}{2\widehat{k}}.
\end{align}

Finally, we summarize the detailed procedures of the BRPSS scheme in \textbf{Algorithm~\ref{BRPSS}}.

\textbf{Remark 2:} The overheads, computational complexity, and hardware requirements of the proposed BRPSS scheme are remarked here. Since the BRPSS is developed based on the phase shifts of the subarrays, only one pilot is needed for beam refinement benefiting from the subarray architecture of the partially-connected hybrid combining structure. The proposed BRPSS has closed-form expressions, which no longer needs running any algorithms and therefore has very low computational complexity. The computation of the BRPSS mainly comes from the computing of \eqref{FoD}, \eqref{SoD}, \eqref{widehatk} and \eqref{widehatb}, which are all only related to $N_{\rm RF}$. Therefore, the computational complexity of the BRPSS is $\mathcal{O}(N_{\rm RF})$.  Note that the unwrapped phase of $\widehat{z}_t$ is the quadratic function of the indices of subarrays. Therefore, at least three subarrays are needed for the BRPSS scheme.

\section{Near-Field Beam Tracking}\label{NFBT}
In this section, we propose a low-complexity NFBT scheme, where a kinematic model is adopted to characterize the channel variations at different time instants and the BRPSS scheme is used to estimate the real-time channel parameters. Then the kinematic model and real-time estimates are exploited by the EKF to track and predict the near-field channel parameters.

\begin{figure}[!t]
	\begin{center}
		\includegraphics[width=70mm]{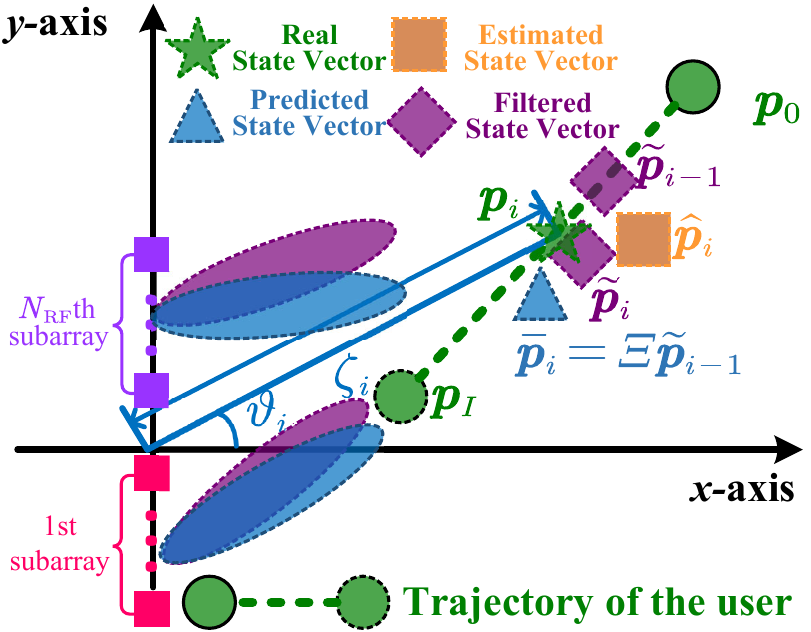}
	\end{center}
	\caption{Illustration of the near-field beam tracking procedure.}
	\label{TrackPlot}
	\vspace{-0.3cm}
\end{figure}

Different from the far-field channel that only depends on  the angle of the user, the near-field channel is related to both the angle and  the distance of the user. By performing the BRPSS scheme with the channel measurements, we can estimate the angles and distances of the user. Then, based on the estimated parameters, the real-time position of the user can be calculated. Given this, we adopt the kinematic model to characterize the channel variations at different time instants. In Fig.~\ref{TrackPlot}, we illustrate the near-field beam tracking procedure, where the center of the BS antennas is set as the origin, the normal direction of the BS antenna array is set as the x-axis and the direction along the BS antennas array is set as the y-axis. We assume the user  communicates with the BS in a block-wise way, where the duration of one block is $\Delta T$. The user moves along the trajectory for $I$ blocks. Denote the kinematic parameters of the user at the $i$th block as $\boldsymbol{p}[i] = [a_x[i],a_y[i],v_x[i],v_y[i]]^{\rm T}$, for $i=0,1,2,\cdots,I$, where $a_x[i]$, $a_y[i]$, $v_x[i]$, and $v_y[i]$ are the x-axis coordinate, the y-axis coordinate, the velocity component in the x-axis direction, and the velocity component in the y-axis direction, respectively. The angle and distance of the user at the $i$th block are expressed as $\vartheta_i$ and $\zeta_i$, respectively. We then propose the NFBT scheme, which includes initialization, beam prediction, beam refinement, filtering, and stop conditions.

\subsubsection{Initialization}
We denote the initial kinematic parameters of the user as $\boldsymbol{p}[0] = [a_x[0],a_y[0],0,0]^{\rm T}$, which can be obtained via the THBT and BRPSS schemes. 
\subsubsection{Beam Prediction}\label{NFBT_BP}
Denote the filtered state vector after performing the EKF at the $(i-1)$th block for $i\ge 1$ as $\widetilde{\boldsymbol{p}}[i-1]$, where  $\widetilde{\boldsymbol{p}}[0] = \boldsymbol{p}[0]$. Then the predicted state vector at the $i$th block can be expressed as 
\begin{align}\label{BeamPredict}
	&\overline{\boldsymbol{p}}[i] = \underset{\boldsymbol{\varXi}}{\underbrace{\left[ \begin{matrix}
		1&		0&		\Delta T&		0\\
		0&		1&		0&		\Delta T\\
		0&		0&		1&		0\\
		0&		0&		0&		1\\
	\end{matrix} \right]}} \widetilde{\boldsymbol{p}}[i-1]
\end{align}
where the predicted distance and angle can be respectively expressed as
\begin{align}\label{PredictedDA}
	\overline{\zeta}_i = \sqrt{\overline{a}_x[i]^2 + \overline{a}_y[i]^2}~\mbox{and}~\overline{\vartheta}_i = \arctan(\overline{a}_y[i]/\overline{a}_x[i]).
\end{align}	

\begin{algorithm}[!t]
	\caption{Near-Field Beam Tracking (NFBT) Scheme}
	\label{algNFBT}
		\begin{algorithmic}[1]
			\STATE \textbf{Input:}$N$, $N_{\rm RF}$, $M$, $\lambda$, $\boldsymbol{p}[0]$ and $\Delta T$.
			\STATE \textbf{Initialization:} $i\leftarrow 1$.
			\WHILE{stop conditions in Sec.~\ref{NFBT_SC} are not satisfied}
			\STATE Obtain $\overline{\boldsymbol{p}}[i]$ via \eqref{BeamPredict}. 
			\STATE Obtain $\widehat{\boldsymbol{p}}[i]$ via \eqref{BeamRefineR}.
			\STATE Obtain $\widetilde{\boldsymbol{p}}[i]$ via the EKF.
			\STATE Obtain $\widetilde{\boldsymbol{h}}_i$ via \eqref{BeamFilterh}.
			\STATE $i\leftarrow i + 1$.			
			\ENDWHILE
			\STATE \textbf{Output:} $\widetilde{\boldsymbol{h}}_i$.
		\end{algorithmic}
\end{algorithm}

\subsubsection{Beam Refinement}\label{NFBT_BR}
In the stage of beam refinement, the user transmits one uplink pilot and the BS receives the signal with the analog combiner
\begin{equation}
\overrightarrow{\boldsymbol{W}}_i = {\rm blkdiag}\{\overrightarrow{\boldsymbol{w}}_1^{\rm H},\overrightarrow{\boldsymbol{w}}_2^{\rm H},\ldots,\overrightarrow{\boldsymbol{w}}_{N_{\rm RF}}^{\rm H}\},
\end{equation}
where 
\begin{align}\label{TempW}
	&[\overrightarrow{\boldsymbol{w}}_t]_m = e^{j\pi(\overline{k}_i((t-1)M + m)^2 + \overline{b}_i((t-1)M + m)}
\end{align}
for $m =1,\ldots,M$ and $t=1,2,\ldots,N_{\rm RF}$. In \eqref{TempW},
\begin{align}
	&\overline{k}_i = -\frac{\lambda(1-\sin^2\overline{\vartheta}_i)}{4\overline{\zeta}_i}~\mbox{and}~\overline{b}_i = -\sin\overline{\vartheta}_i - \overline{k}_i(N+1)
\end{align}
which are calculated based on the predicted distance and angle in \eqref{PredictedDA}. Then the received signal of the $t$th subarray after analog combining can be expressed as 
\begin{align}
\overline{z}_t = \overrightarrow{\boldsymbol{w}}_t^{\rm H}\boldsymbol{G}_t\overline{\boldsymbol{h}}_i + \overrightarrow{\boldsymbol{w}}_t^{\rm H}\boldsymbol{G}_t\overline{\boldsymbol{\eta}}_i,
\end{align}
for $t=1,2,\ldots,N_{\rm RF}$, where $\overline{\boldsymbol{h}}_i$ denotes the channel between the user and the BS at the $i$th block and $\overline{\boldsymbol{\eta}}_i$ denotes the AWGN. Substituting $N$, $N_{\rm RF}$, $M$, $\lambda$, $\overline{z}_t$, $-\sin\overline{\vartheta}_i$ and $\overline{\zeta}_i$ into \textbf{Algorithm~\ref{BRPSS}}, we can obtain the estimated angle $\widehat{\vartheta}_i$ and the estimated distance $\widehat{\zeta}_i$ of the user at the $i$th block. Then the estimated state vector can be expressed as 
\begin{align}\label{BeamRefineR}
	\widehat{\boldsymbol{p}}[i] = \big[\widehat{\zeta}_i\cos\widehat{\vartheta}_i,\widehat{\zeta}_i\sin\widehat{\vartheta}_i,0,0\big]^{\rm T}.
\end{align}

\subsubsection{Filtering}\label{NFBT_F}
The estimated stated vector, $\widehat{\boldsymbol{p}}[i]$, and the predicted state vector, $\overline{\boldsymbol{p}}[i]$, can be  exploited by the widely-adopted EKF for effective tracking. Since the EKF has been exhaustively introduced in a variety of works~\cite{FSSP,JSAC22Liufan,TWC20Yuanweijie}, we omit the details and denote the filtered state vector at the $i$th block after performing the EKF as $\widetilde{\boldsymbol{p}}[i]$. Then the estimated channel steering vector of the LoS path at the $i$th block can be expressed as 
\begin{align}\label{BeamFilterh}
 &[\widetilde{\boldsymbol{h}}_i]_n = \frac{1}{\sqrt{N}}e^{j\pi\big(\overrightarrow{k}_i n^2 + \overrightarrow{b}_i n\big)} \nonumber\\
 &\overrightarrow{k}_i = -\frac{\lambda(1-\sin^2\widetilde{\vartheta}_i)}{4\widetilde{\zeta}_i},~\overrightarrow{b}_i = -\sin\widetilde{\vartheta}_i - \overrightarrow{k}_i(N+1), \nonumber \\
 &\widetilde{\zeta}_i = \sqrt{\widetilde{a}_x[i]^2 + \widetilde{a}_y[i]^2},~\mbox{and}~\widetilde{\vartheta}_i = \arctan(\widetilde{a}_y[i]/\widetilde{a}_x[i]),
\end{align}
for $n=1,2,\ldots,N$, which can be used to design the hybrid combiners for communication. 
\subsubsection{Stop Condition}\label{NFBT_SC}
We repeat Section~\ref{NFBT_BP}, Section~\ref{NFBT_BR}, and Section~\ref{NFBT_F} until the predefined maximum number of blocks $I$ is achieved or the communication process is completed.

In Fig.~\ref{TrackPlot}, we illustrate the process of beam training using the $i$th block as an  example. Suppose we have obtained the filtered state vector in the $(i-1)$th block, $\widetilde{\boldsymbol{p}}[i-1]$. Based on the  kinematic model in \eqref{BeamPredict}, we can obtain the predicted state vector in the $i$th block, $\overline{\boldsymbol{p}}[i]$.  Using $\overline{\boldsymbol{p}}[i]$ as a reference, the BS designs the analog combiner $\overrightarrow{\boldsymbol{W}}_i$ and each subarray independently receives the uplink pilot  with $\overrightarrow{\boldsymbol{W}}_i$. Based on the measurements of subarrays, the BS performs the BRPSS scheme to obtain the estimated state vector, $\widehat{\boldsymbol{p}}[i]$. Then, by exploiting the EKF, the BS fuses $\overline{\boldsymbol{p}}[i]$ and  $\widehat{\boldsymbol{p}}[i]$ to obtain the more accurate filtered state vector~$\widetilde{\boldsymbol{p}}[i]$. During the process of beam tracking, for each block, the BS only receives one uplink pilot to obtain the estimated state vector. Therefore, the overhead of the proposed NFBT scheme is one.

Finally, we summarize the detailed procedures of the proposed NFBT scheme in \textbf{Algorithm~\ref{algNFBT}}.

\textbf{Remark 3:} The proposed schemes are adaptable to the multi-user scenarios with three extensions. Firstly, following the works in \cite{Tcom22CMH}, we can allocate mutually orthogonal pilots to  $N_{\rm UE}$ users and separately perform the proposed schemes for each user. Secondly, following the works in \cite{TWC17XZY},  we allocate mutually orthogonal pilots to the $N_{\rm RF}$ RF chains and perform the downlink beam training/tracking. Then, the users can concurrently  perform the proposed schemes. Thirdly, following the works in \cite{TWC20NS}, we perform uplink beam training for the $N_{\rm UE}$ users simultaneously, where $\widehat{L}$ channel paths are detected sequentially based on the  beam training results. Then, the users transmit orthogonal pilots to determine the association of the $\widehat{L}$ detected paths.

\begin{table*}[!t]
	\centering
	\renewcommand{\arraystretch}{1.1}
	\caption{Overheads for different beam training schemes.}\label{tab1}
	
		\begin{tabular}{cccc}
			
			\hline\hline\noalign{\smallskip}	
			\textbf{Schemes} &  \textbf{Overhead}  &  \textbf{Parameter Settings} & \textbf{Calculated Training Overhead }\\
			\noalign{\smallskip}\hline\noalign{\smallskip}
			HFBS & $Q(S+1)$ & $Q = 512, S = 11$  & 6144  \\
			FFBT~\cite{TWC15AAl} & $Q$& $Q = 512$  & 512  \\
			TPBT~\cite{WCL22ZYP} & $Q + K(S+1)$ & $Q = 512,K = 3, S=11$ & 548  \\	
			DHBT~\cite{CC22WXH} & $VR$ & $V = 2,R = 256$ & 512   \\
			P-SOMP~\cite{Tcom22CMH} & $P$ & $P=128$ & 128    \\
			Proposed THBT & $M$ & $M=128$ & 128 \\
			Proposed THBT with BRPSS & $M+1$ & $M=128$ & 129 \\
			\noalign{\smallskip}\hline
	\end{tabular}
	\vspace{-0.5cm}
\end{table*}


\section{Simulation Results}\label{SimulationResults}
Now we evaluate the performance of the proposed schemes. We consider an XL-MIMO system equipped with $N = 512$ antennas~\cite{JSAC23WZD}. The antenna array is composed of $N_{\rm RF} = 4$ subarrays with each subarray having $M = 128$ antennas~\cite{TWC20SXS}. The wavelength is set to be $\lambda = 0.003{\rm~m}$ corresponding to the carrier frequency of 100${\rm~GHz}$~\cite{Tcom22CMH}. The channel between the user and the BS is set up with $L = 3$ channel paths with one LoS path and two NLoS paths, where the channel gain of the LoS path obeys $g_1\sim\mathcal{CN}(0,1)$ and the NLoS paths obey  $g_2\sim\mathcal{CN}(0,0.01)$ and $g_3\sim\mathcal{CN}(0,0.01)$~\cite{TWC20QCH}. Channel angle $\Omega_l$ of the $l$th path obeys the uniform distribution between $[-\sqrt{3}/2,\sqrt{3}/2]$~\cite{Tcom22CMH}. The HFBS, far-field beam sweeping (FFBS)~\cite{TWC15AAl}, TPBT~\cite{WCL22ZYP}, DHBT~\cite{CC22WXH} and P-SOMP~\cite{Tcom22CMH} are adopted as the benchmarks. The parameters of different schemes are set in the \textbf{Parameter Settings} of Table~\ref{tab1}, where $K$, $V$ and $R$ denote the number of candidate angles, the number of layers in the hierarchical codebook and the number of space samples in each layer.

\begin{figure}[!t]
	\begin{center}
		\includegraphics[width=68mm]{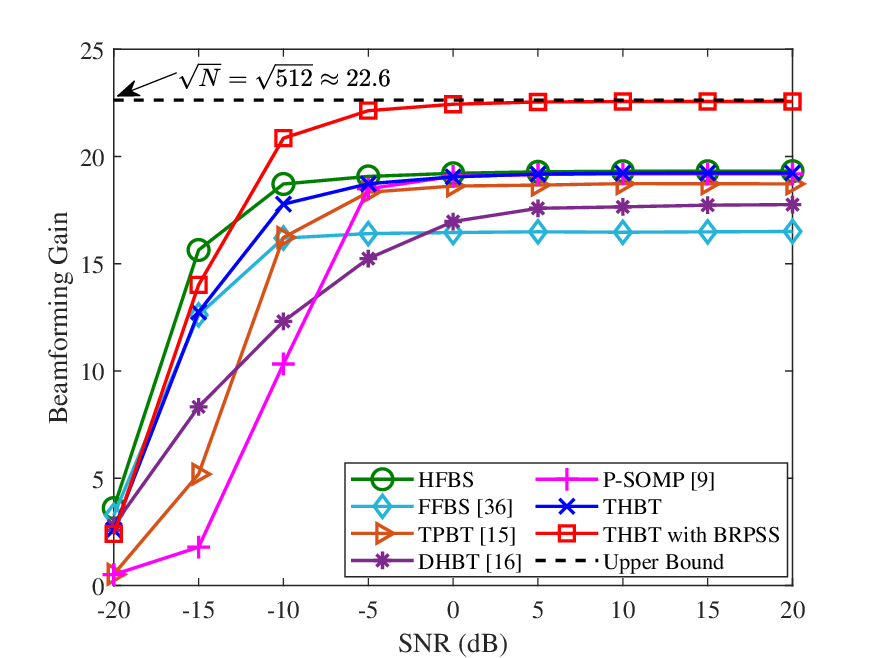}
	\end{center}
	\caption{Comparisons of beamforming gains for different schemes.}\label{fig:SE}
		\vspace{-0.2cm}
\end{figure}

\subsection{Evaluation of the THBT and BRPSS Schemes}
In Fig.~\ref{fig:SE}, we compare the proposed schemes with the HFBS, FFBS, TPBT, DHBT and P-SOMP in terms of beamforming gains for different SNRs. Suppose $\boldsymbol{f}$ is the estimated channel steering vector. Then, the beamforming gain after beam alignment is defined as 
\begin{align}
	\xi  = \max_l\frac{|g_l|}{g_{\rm m}} |\boldsymbol{\alpha}(N,\Omega_l,r_l)^{\rm H}\boldsymbol{f}|,
\end{align} 
where $g_{\rm m} = \max_l |g_l|$. The distances between the BS and the user or scatterers obey the uniform distribution between $[6,150]$~${\rm m}$, where the lower bound is set according to the reactive distance~\cite{Tcom22CMH} while the upper bound is set according to the effective  Rayleigh distance~\cite{Arxiv21_CMY}. The hybrid-field codebook is designed with $Q = 512$ and $S = 11$, which is also used to generate the dictionary of the P-SOMP for fair comparison. From Fig.~\ref{fig:SE}, the HFBS can achieve better performance than the other schemes when the SNR is less than $-10$~dB, which lies in the fact that the HFBS exhaustively tests all the codewords in $\boldsymbol{C}_{\rm h}$ and needs far more times of beam training than the other schemes. In addition, the proposed THBT with the BRPSS scheme performs the best among all the schemes when the SNR is larger than $-10$~dB because the performance of all the other schemes is limited by the quantized error of the codebook or dictionary while the proposed THBT with the BRPSS scheme can achieve high-resolution estimation thanks to the beam refinement of the BRPSS. Moreover, the performance of the proposed THBT scheme can approach the performance of the HFBS at various SNR conditions. The performance of P-SOMP is worse than that of the other schemes at low SNRs, such as -15~dB, because the random beamforming of P-SOMP cannot achieve enough beamforming gain and will significantly degrade the performance. The performance of the FFBS is worse than that of the other schemes at high SNRs because the beamforming gain of the far-field channel steering vector will decrease in the near field. Furthermore, the TPBT performs worse than the proposed THBT; the justification is that the first phase of TPBT may select the incorrect middle-$K$ angles due to the effects of the noise and the NLoS paths. The DHBT performs worse than the proposed THBT due to the neglect of the polar-domain sparsity.

\begin{figure}[!t]
	\begin{center}
		\includegraphics[width=68mm]{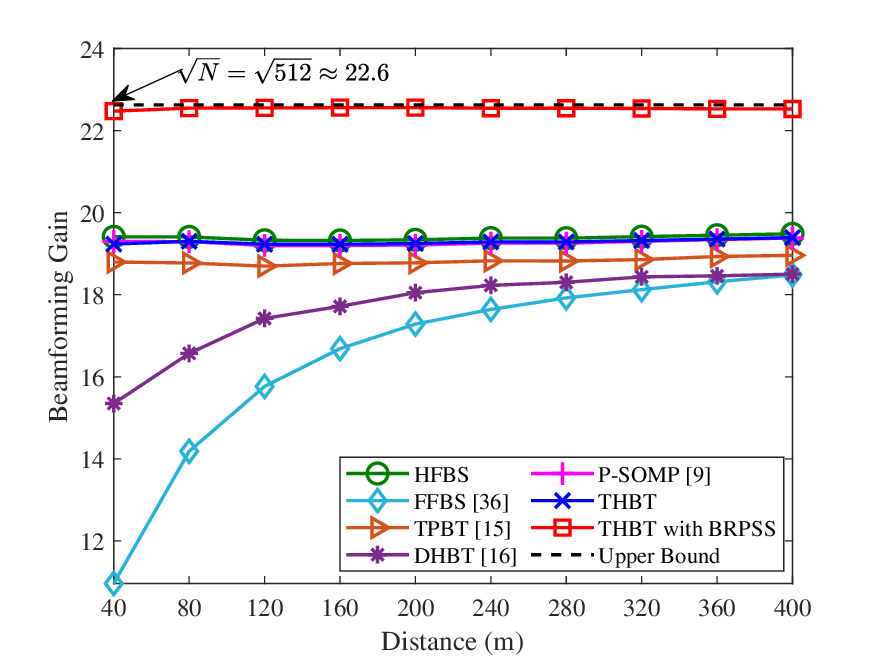}
	\end{center}
	\caption{Comparisons of the beamforming gains for different schemes with varying distances.}\label{fig:BG}
		\vspace{-0.2cm}
\end{figure}

In Fig.~\ref{fig:BG}, we compare the proposed schemes with the HFBS, FFBS, TPBT, DHBT and P-SOMP in terms of beamforming gains for different distances. The distances between the BS and the user or scatterers obey the uniform distribution between $[6,r]{\rm~m}$, where $r$ ranges from $40$ to $400$. Note that $400$~m is approximately the Rayleigh distance of the considered XL-MIMO system. The SNR is fixed to be $10$~dB. From Fig.~\ref{fig:BG}, the proposed THBT with the BRPSS scheme achieves the highest beam gain at different distances thanks to the high-resolution BRPSS scheme. As the distance decreases, the FFBS will suffer from severe loss of beamforming gain because it only considers the far-field channels. In addition, the performance of DHBT deteriorates with the decrease of the distance as it neglects the  polar-domain sparsity, which makes it sensitive to the near-field effects. By contrast, both the THBT scheme and the THBT with BRPSS scheme are robust to the distance, which indicates that  the proposed beam training and beam refinement schemes are adaptive to both the near field and the far field.

\begin{figure}[!t]
	\begin{center}
		\includegraphics[width=68mm]{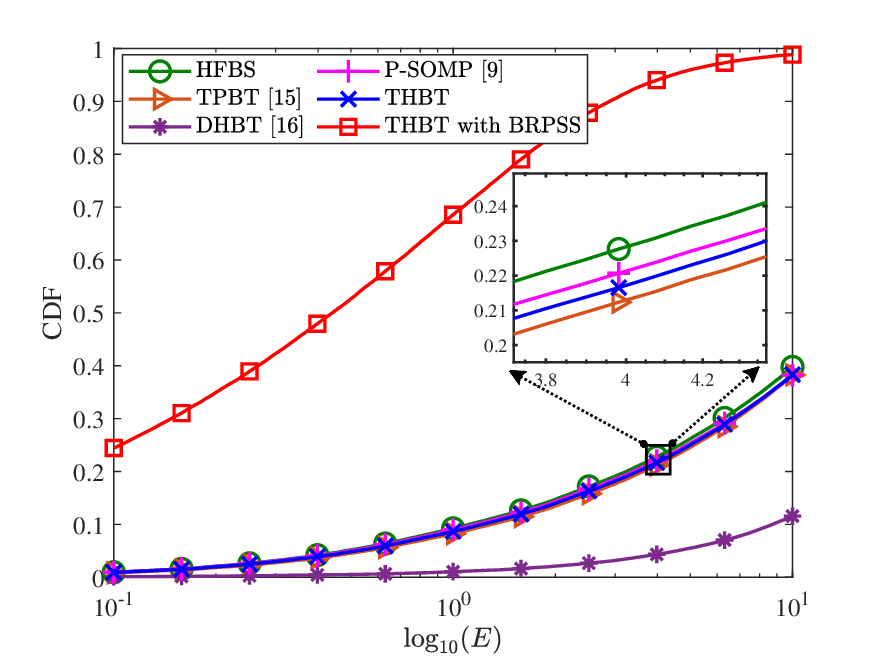}
	\end{center}
	\caption{Evaluation of the positioning  performance for different schemes.}\label{fig:PE}
\end{figure}

\begin{figure}[!t]
	\begin{center}
		\includegraphics[width=68mm]{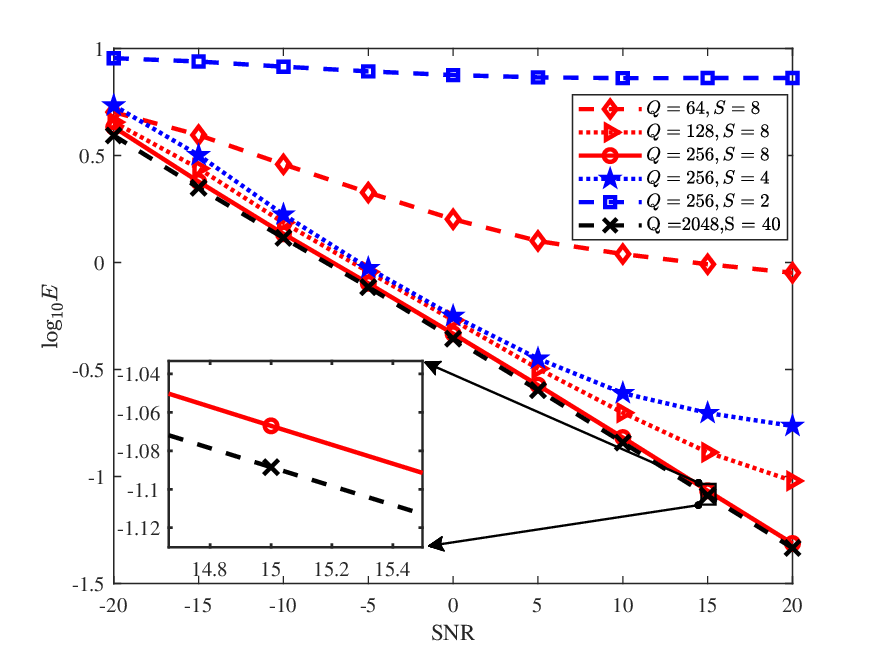}
	\end{center}
	\caption{Evaluation of the BRPSS for different settings of $Q$ and $S$.}
	\label{VeriBRPSS}
	\vspace{-0.25cm}
\end{figure}

In Fig.~\ref{fig:PE}, we evaluate the positioning performance of the proposed schemes, where the SNR is set to be $20$~dB. The distances between the BS and the user or scatterers obey the uniform distribution between $[6,150]$~m, which is the same as in Fig.~\ref{fig:SE}.  Since the FFBS cannot obtain the position of the user, we only compare the proposed schemes with HFBS, TPBT, DHBT and P-SOMP.  We denote the distance between the real position and the estimated position as $E$ and use $10^5$ times of Monte Carlo simulation to calculate the cumulative distribution function (CDF).  From Fig.~\ref{fig:PE}, the proposed THBT with the BRPSS scheme has the smallest positioning error  among all the schemes due to the high-resolution estimation of the BRPSS scheme. In addition, the HFBS, TPBT, THBT and P-SOMP share similar performance because they are based on the same quantized angles and distances. The DHBT performs worse than the other schemes due to the neglect of the polar-domain sparsity.

In Table.~\ref{tab1}, we compare the training overheads of different schemes. The training overheads of the HFBS, FFBS, TPBT, DHBT, P-SOMP, the proposed THBT, and the proposed THBT with the BRPSS are $Q(S+1)$, $Q$, $Q+K(S+1)$, $VR$, $P$, $M$ and $M+1$, respectively. Under the simulation settings, these seven schemes require $6144$, $512$, $548$, $512$, $128$, $128$, and $129$ times of beam training, respectively.  Specifically, the performance of the proposed THBT scheme can approach that of the HFBS with $97.92\%$ reduction in training overhead, and the proposed THBT with the BRPSS scheme outperforms the HFBS with $97.90\%$ reduction in training overhead for high~SNRs.

\begin{figure}[!t]
	\begin{center}
		\includegraphics[width=68mm]{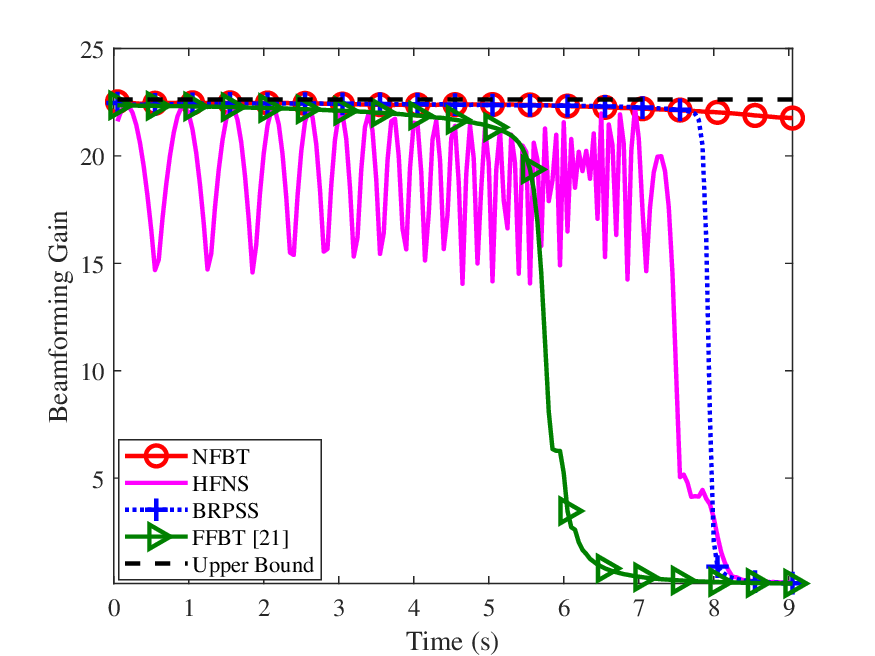}
	\end{center}
	\caption{Comparisons of the beamforming gains during the tracking process.}\label{figBTRT}
\end{figure}
\begin{figure}[!t]
	\begin{center}
		\includegraphics[width=68mm]{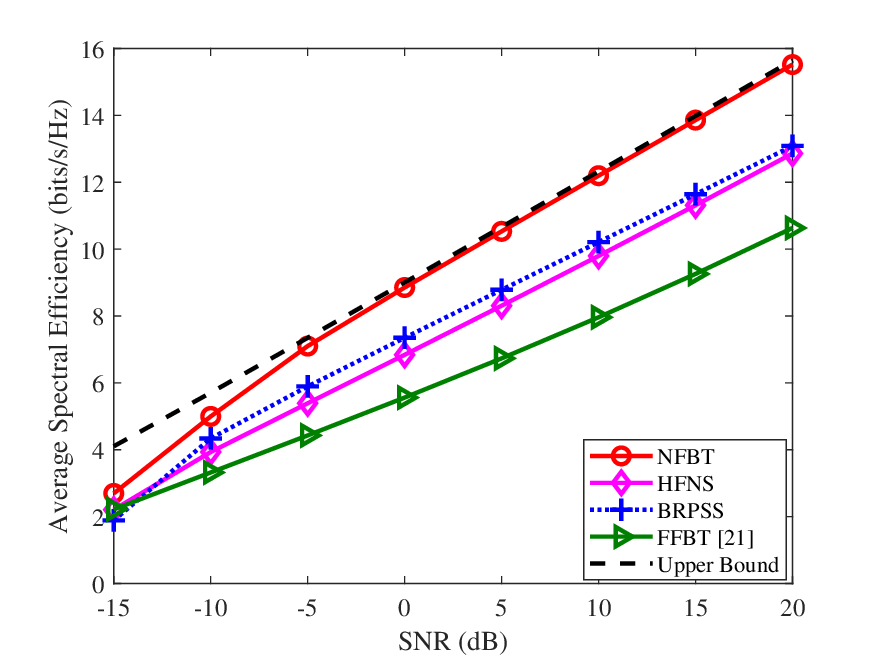}
	\end{center}
	\caption{Comparisons of the spectral efficiency during the tracking process.}\label{figBTSE}
	\vspace{-0.1cm}
\end{figure}

In Fig.~\ref{VeriBRPSS}, we evaluate the performance of the BRPSS for different settings of $Q$ and $S$.  To avoid the effects of the NLoS paths and the prior beam training failure, we set $L = 1$ and suppose that the  beam training has found the codeword best fitting for the channel in the hybrid-field codebook. The distances between the BS and the user obey the uniform distribution between $[10,30]$~m. From~Fig.~\ref{VeriBRPSS}, if we fix $Q = 256$ or $S=8$, the positioning error decreases with the increase of $S$ or $Q$. This is because  beam gain losses of subarrays will occur if the conditions of $Q$ and $S$ in \eqref{MinS} are not satisfied, which will deteriorate the performance of beam refinement. However, if the conditions of $Q$ and $S$ in \eqref{MinS} are satisfied, i.e. $Q = 256$ and $S = 8$, continuing to increase $Q$ and $S$ can only slightly improve the performance of beam refinement, which verifies that \eqref{MinS} is appropriate for the settings of $Q$ and $S$.

\subsection{Evaluation of the NFBT Scheme}
Now we evaluate the performance of the proposed NFBT scheme. We set $\boldsymbol{p}_0 = [50,50\sqrt{3},0,0]$, $\Delta T = 0.05$~s, $I = 180$,  and $[v_x[i],v_y[i]]^{\rm T} = [-5,-5\sqrt{3}]^{\rm T}$  for $i =1,2,\ldots,I$.

In Fig.~\ref{figBTRT}, we evaluate the tracking performance of different schemes in terms of  the beamforming gains during the tracking process. We extend the neighboring search in \cite{TWC22NBY} to the hybrid-field neighboring search (HFNS), which tests the  additional four codewords adjacent to the best codeword of the previous block.  The  BRPSS and the far-field beam tracking (FFBT)~\cite{TWC18ZDL} are also adopted as benchmarks, where the BRPSS is performed  based on the estimation of the previous block. The SNR is set to be $0$~dB. From the figure, the performance of the HFNS drops dramatically at the $7.5$th~second and that of the BRPSS drops dramatically at the $8$th~second because the deviation between the positions of the adjacent blocks is too large and the deteriorated beamforming gains cannot support effective beam tracking. In addition, the performance of the FFBT drops dramatically at the $6$th second because  the near-field channel  differs gradually from the far-field channel with the decrease of the distance.  However, the proposed NFBT scheme can maintain high beamforming gains during the whole process of beam tracking. It is also worth noting that the performance of the proposed NFBT scheme decreases slightly with the increase of time. This is because the near-field effect strengthens gradually with the decrease of distance and the same tracking error will lead to larger beamforming gain loss for smaller distances. 

\begin{table}[!t]  
	\centering
	\renewcommand{\arraystretch}{1.0}
	\caption{Overheads for different beam tracking schemes.}\label{tab2}
	\begin{tabular}{cc}
		\hline\hline\noalign{\smallskip}	
		\textbf{Schemes} &  \textbf{Overhead}\\
		\noalign{\smallskip}\hline\noalign{\smallskip}
		NFBT  &   1  \\
		HFNS  &   5 \\
		BRPSS &   1 \\
		FFBT~\cite{TWC18ZDL}  &  3   \\
		\noalign{\smallskip}\hline
	\end{tabular}
	\vspace{-0.1cm}
\end{table}

In Fig.~\ref{figBTSE}, we compare the proposed NFBT with the HFNS, BRPSS and FFBT in terms of the average spectral efficiency during the tracking process for different SNRs. From the figure, the proposed NFBT performs the best among the four schemes and can approach the upper bound when SNR is larger than $-5$dB, which demonstrates the effectiveness of the proposed schemes.

In Table~\ref{tab2}, we also compare the overheads of different beam tracking schemes. Since the HFNS tests the best codeword  of the previous block and additional four codewords adjacent to the best codeword, it needs five times of beam training. In addition, the FFBT tests the best codeword  of the previous block and the additional two codewords adjacent to the best codeword, and it needs three times of beam training. According to Section~\ref{SecBRPSS}, the overhead of the BRPSS scheme is one. Similarly, the overhead of the NFBT scheme, which is based on the BRPSS scheme, is also one. In summary, the proposed NFBT scheme outperforms existing ones with lower overhead.

\section{Conclusion}\label{Conclusion}
In this paper, a THBT scheme has been proposed. In the first stage, each subarray independently uses multiple far-field channel steering vectors to approximate near-field channel steering vectors for analog combining. In the second stage, digital combiners are designed to combine the outputs of the analog combiners from the first stage to find the codeword in the predefined hybrid-field codebook best fitting for the channel. Then, based on the PSP and the time-frequency duality, the expressions of subarray signals after analog combining have been analytically derived, and the BRPSS scheme with closed-form solutions has been proposed for high-resolution channel parameter estimation. Moreover, a low-complexity NFBT scheme has been proposed, where the kinematic model has been adopted to characterize the channel variations at different time instants and the EKF has been exploited for beam tracking. For future works, we will try to perform effective beam refinement for XL-MIMO systems with the power of received signals.

\section*{Appendix A}
First of all, we calculate the maximum received powers with the near-field and the far-field steering vectors, respectively. Suppose the $t$th subarray receives signals with far-field steering vector $\sqrt{M}\boldsymbol{\beta}(M,\Omega)$. Then, the absolute value of the received signal after analog combining can be calculated as 
\begin{align}\label{NFSA}
	|G(\boldsymbol{G}_t\boldsymbol{h},\Omega)| &= \sqrt{M}\boldsymbol{\beta}(M,\Omega)^{\rm H}\boldsymbol{G}_t\boldsymbol{h}      \nonumber \\
	&\approx \left|\sum_{n=(t-1)M +1}^{tM} e^{j\pi(kn^2+ bn)}e^{-j\pi n\Omega}\right| \nonumber \\
	&= \left|\sum_{n = 1}^{M} e^{j\pi(kn^2+ \overline{\overline{b}}n)}e^{-j\pi n\Omega}\right| \nonumber \\
	&\approx \left|\int_{1}^{M} e^{j\pi(k\chi^2+ (\overline{\overline{b}}-\Omega)\chi))} {\rm d}\chi\right| \nonumber \\
	&= \left|\int_{-\infty}^{\infty} U(\chi)e^{j\pi(k\chi^2+ (\overline{\overline{b}}-\Omega)\chi))} {\rm d}\chi\right|,
\end{align}
where
\begin{align}\label{Uz}
	&\overline{\overline{b}} \triangleq b + 2k(t-1)M,~~U(\chi) = \left\{ \begin{array}{ll}
		1, & 1\leq \chi \leq M,\\
		0, & \mbox{others}.
	\end{array} \right.
\end{align}
According to the PSP~\cite{AMFI,AMMSE,ASF}, \eqref{NFSA} can be approximated as
\begin{align}\label{PSPApp}
	|G(\boldsymbol{G}_t\boldsymbol{h},\Omega)| &\approx \left|\sqrt{\frac{-2\pi}{\Phi''(\chi_0,\Omega)}}U(\chi_0)\right| \nonumber\\
	&=\left\{ \begin{array}{ll}
		\!\sqrt{\frac{1}{-k}},\!&\!\overline{\overline{b}}\!+\!2kM\!\leq\!\Omega\!\leq\!\overline{\overline{b}}\!+\!2k,\\
		0, &\!\mbox{others},
	\end{array} \right.
\end{align}
where $\Phi(\chi,\Omega) \triangleq  \pi(k\chi^2+ (\overline{\overline{b}}-\Omega)\chi))$ and $\chi_0 = \frac{\Omega - \overline{\overline{b}}}{2k}$. The relation in \eqref{PSPApp} indicates that the maximum received powers with near-field channel steering vectors can be approximated as $\sqrt{\frac{1}{-k}}$.  On the other hand, when the analog combiner is aligned with the channel steering vector, the modulus of the maximum received signal after analog combining is $M$. Therefore the beam gain loss of replacing the near-field channel steering vector with the far-field channel steering vector can be approximated as
\begin{align}
	\Gamma &= 1 - \frac{\max_{\Omega}|G(\boldsymbol{G}_t\boldsymbol{h},\Omega)|}{M} \nonumber \\
	&=1 - \sqrt{\frac{4r}{\lambda (1-\Omega^2) M^2}}\nonumber \\
	&\overset{\rm (a)}{\le} 1 - \sqrt{\frac{4r}{\lambda M^2}}\nonumber \\
	&\overset{\rm (b)}{\le} 1 - \sqrt{\frac{2\sqrt{D^3/\lambda}}{\lambda M^2}}\nonumber \\
	&=1-\frac{N_{\rm RF}}{\sqrt[4]{{2N}}},
\end{align}
where the equation in $\rm (a)$ holds when $\Omega = 0$ and the equation in $\rm (b)$ holds when $r = 0.5\sqrt{D^3/\lambda}$. Since the beam gain loss cannot be less than $0$, we normalize the maximum beam gain loss as 
\begin{align}
	\Gamma_{\rm max} = \max\{{1-{N_{\rm RF}}/{\sqrt[4]{{2N}}}},0\},
\end{align}
which completes the proof.
\section*{Appendix B}
First of all, we calculate the beam center of $\boldsymbol{G}_t\boldsymbol{\alpha}(N,\Omega,r)$. Based on \eqref{PSPApp},  we can approximate the beam gain of $\boldsymbol{G}_t\boldsymbol{\alpha}(N,\Omega,r)$  as 
\begin{align}\label{PSPApp2}
	|G(\boldsymbol{G}_t\boldsymbol{\alpha}(N,\Omega,r),\Omega)| =\left\{ \begin{array}{ll}
		\!\sqrt{\frac{1}{-k}},\!&\!\overline{\overline{b}}\!+\!2kM\!\leq\!\Omega\!\leq\!\overline{\overline{b}}\!+\!2k,\\
		0, &\!\mbox{others},
	\end{array} \right.
\end{align}
which indicates that the beam coverage of $\boldsymbol{G}_t\boldsymbol{\alpha}(N,\Omega,r)$ is $[\overline{\overline{b}}+2kM,\overline{\overline{b}}+2k]$. Therefore, the beam center of $\boldsymbol{G}_t\boldsymbol{\alpha}(N,\Omega,r)$ can be calculated as 
\begin{align}\label{Bc}
	B_{t} &= \overline{\overline{b}} + k(M+1) \nonumber \\
		      &= \Omega + \frac{\lambda(1-\Omega^2)(N-(2t-1)M)}{4r}.
\end{align}
Then we calculate  $\Psi_t$, which denotes the sine result of the angle that points from the center of the $t$th subarray to the user. As shown in Fig. \ref{MultipathChannelModel}, the center of the $t$th subarray is $(0,\Delta_t\lambda)$, where $\Delta_t = [(2t-1)M-N]/4$. Then $\Psi_t$ can be computed as
\begin{align}\label{Eq.hatOmega}
	\Psi_t & = \frac{r\Omega-\Delta_t\lambda}{\sqrt{r^2+\Delta_t^2\lambda^2-2r\Omega\Delta_t\lambda}} \nonumber \\
	& =\frac{r\Omega-\Omega^2\Delta_t\lambda}{\sqrt{r^2+\Delta_t^2\lambda^2-2r\Omega\Delta_t\lambda}} + \frac{(\Omega^2-1)\Delta_t\lambda}{\sqrt{r^2+\Delta_t^2\lambda^2-2r\Omega\Delta_t\lambda}}\nonumber \\
	&\overset{\rm (a)}{\approx}  \frac{r\Omega-\Omega^2\Delta_t\lambda}{r-\Omega\Delta_t\lambda} + \frac{(\Omega^2-1)\Delta_t\lambda}{r} \nonumber \\
	& = \Omega + \frac{\lambda(1-\Omega^2)(N-(2t-1)M)}{4r},
\end{align}
where the former of $\rm (a)$ holds because we approximate $\sqrt{r^2+\Delta_t^2\lambda^2-2r\Omega\Delta_t\lambda}$ with its first-order Taylor series and  the latter of $\rm (a)$ holds because we approximate $\sqrt{r^2+\Delta_t^2\lambda^2-2r\Omega\Delta_t\lambda}$ with $r$. We use different approximations in $\rm (a)$ of \eqref{Eq.hatOmega} because the numerator of the former is much larger than that of the latter, which implies that the denominator of the former needs a more accurate approximation than that of the latter. Comparing \eqref{Bc} and \eqref{Eq.hatOmega}, we have 
\begin{align}
	B_t\approx \Psi_t,
\end{align}
which completes the proof.

\bibliographystyle{IEEEtran}
\bibliography{IEEEabrv,IEEEexample}

\begin{IEEEbiography}[{\includegraphics[width=1in,height=1.25in,clip,keepaspectratio]{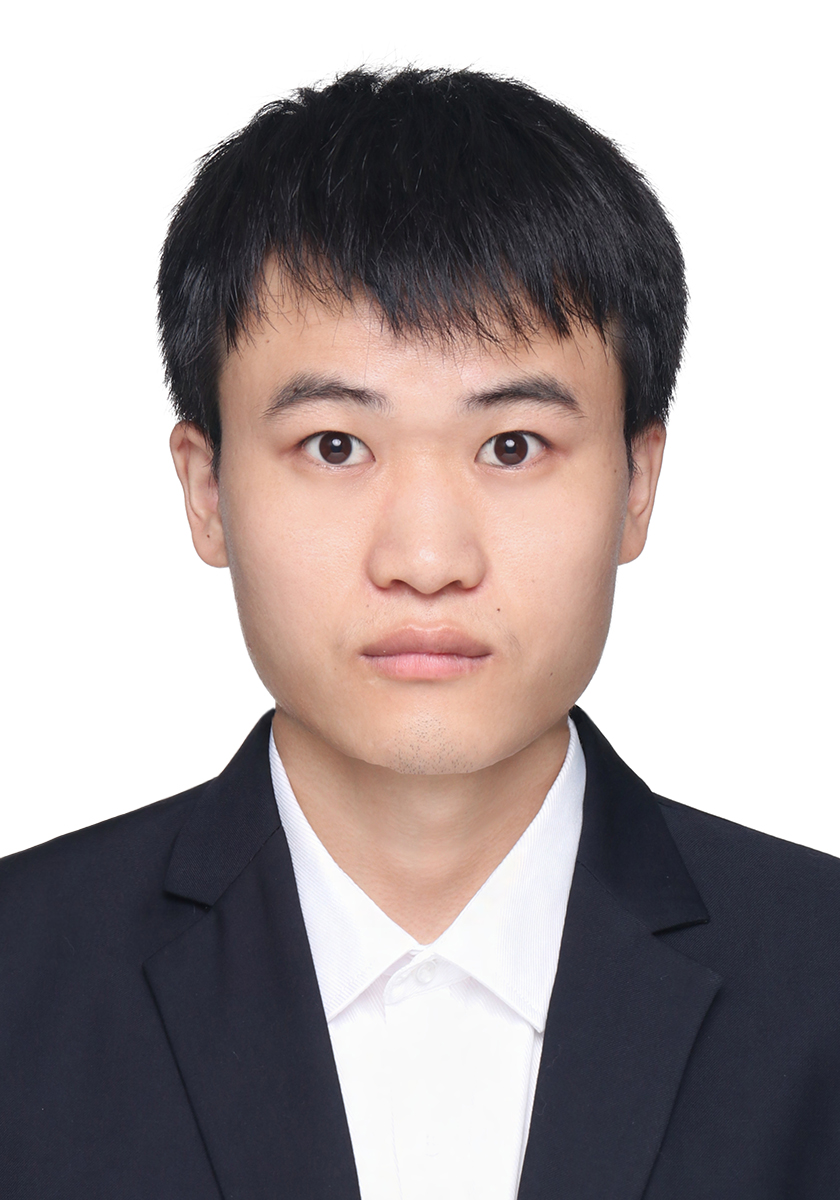}}]{Kangjian Chen}
	(Student Member, IEEE) received the B.S. degree from Nanjing University of Science and Technology, Nanjing, China, in 2017, and the M.S. degree in signal processing from Southeast University, Nanjing, in 2020, where he is currently pursuing the Ph.D. degree. His current research interests include integrated sensing and communication (ISAC), massive MIMO, and reconfigurable intelligent surface (RIS). He received the Best Paper Awards from IEEE Global Communications Conference (GLOBECOM) in 2019 and IEEE/CIC International Conference on Communications in China (ICCC) in~2022.
\end{IEEEbiography}

\begin{IEEEbiography}[{\includegraphics[width=1in,height=1.25in,clip,keepaspectratio]{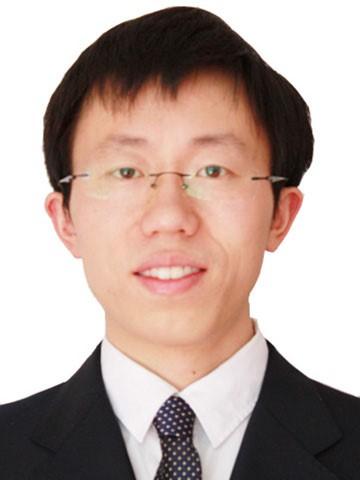}}]{Chenhao Qi}
	(Senior Member, IEEE) received the B.S. degree (Hons.) in information engineering from the Chien-Shiung Wu Honored College, Southeast University, China, in 2004, and the Ph.D. degree in signal and information processing from Southeast University in 2010. 
	
	From 2008 to 2010, he visited the Department of Electrical Engineering, Columbia University, New York, USA. Since 2010, he has been a Faculty Member with the School of Information Science and Engineering, Southeast University, where he is currently a Professor and the Head of Jiangsu Multimedia Communication and Sensing Technology Research Center. He received Best Paper Awards from IEEE GLOBECOM in 2019, IEEE/CIC ICCC in 2022, and the 11th International Conference on Wireless Communications and Signal Processing (WCSP) in 2019. He has served as an Associate Editor for IEEE TRANSACTIONS ON COMMUNICATIONS, IEEE COMMUNICATIONS LETTERS, IEEE OPEN JOURNAL OF THE COMMUNICATIONS SOCIETY, IEEE OPEN JOURNAL OF VEHICULAR TECHNOLOGY, and CHINA COMMUNICATIONS. 
\end{IEEEbiography}

\begin{IEEEbiography}[{\includegraphics[width=1in,height=1.25in,clip,keepaspectratio]{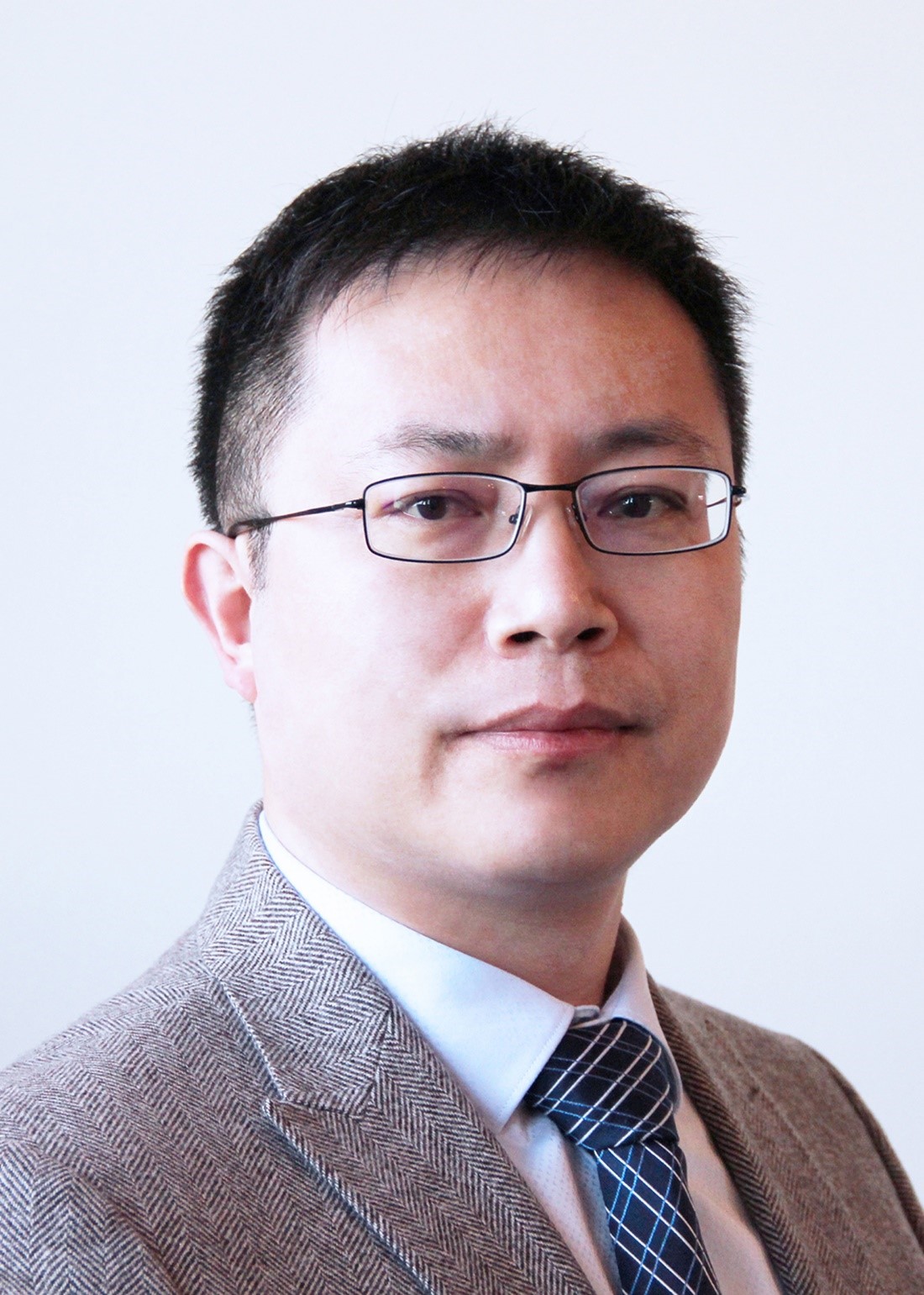}}]{Cheng-Xiang Wang }
	(Fellow, IEEE) received the B.Sc. and M.Eng. degrees in communication and information systems from Shandong University, China, in 1997 and 2000, respectively, and the Ph.D. degree in wireless communications from Aalborg University, Denmark, in 2004.
	
	He was a Research Assistant with the Hamburg University of Technology, Hamburg, Germany, from 2000 to 2001, a Visiting Researcher with Siemens AG Mobile Phones, Munich, Germany, in 2004, and a Research Fellow with the University of Agder, Grimstad, Norway, from 2001 to 2005. He has been with Heriot-Watt University, Edinburgh, U.K., since 2005, where he was promoted to a Professor in 2011. In 2018, he joined Southeast University, Nanjing, China, as a Professor. He is also a part-time Professor with Purple Mountain Laboratories, Nanjing. He has authored 4 books, 3 book chapters, and 520 papers in refereed journals and conference proceedings, including 27 highly cited papers. He has also delivered 24 invited keynote speeches/talks and 16 tutorials in international conferences. His current research interests include wireless channel measurements and modeling, 6G wireless communication networks, and electromagnetic information theory.
	
	Dr. Wang is a Member of the Academia Europaea (The Academy of Europe), a Member of the European Academy of Sciences and Arts (EASA), a Fellow of the Royal Society of Edinburgh (FRSE), IEEE, IET and China Institute of Communications (CIC), an IEEE Communications Society Distinguished Lecturer in 2019 and 2020, a Highly-Cited Researcher recognized by Clarivate Analytics in 2017-2020. He is currently an Executive Editorial Committee Member of the IEEE TRANSACTIONS ON WIRELESS COMMUNICATIONS. He has served as an Editor for over ten international journals, including the IEEE TRANSACTIONS ON WIRELESS COMMUNICATIONS, from 2007 to 2009, the IEEE TRANSACTIONS ON VEHICULAR TECHNOLOGY, from 2011 to 2017, and the IEEE TRANSACTIONS ON COMMUNICATIONS, from 2015 to 2017. He was a Guest Editor of the IEEE JOURNAL ON SELECTED AREAS IN COMMUNICATIONS, Special Issue on Vehicular Communications and Networks (Lead Guest Editor), Special Issue on Spectrum and Energy Efﬁcient Design of Wireless Communication Networks, and Special Issue on Airborne Communication Networks. He was also a Guest Editor for the IEEE TRANSACTIONS ON BIG DATA, Special Issue on Wireless Big Data, and is a Guest Editor for the IEEE TRANSACTIONS ON COGNITIVE COMMUNICATIONS AND NETWORKING, Special Issue on Intelligent Resource Management for 5G and Beyond. He has served as a TPC Member, a TPC Chair, and a General Chair for more than 30 international conferences. He received 15 Best Paper Awards from IEEE GLOBECOM 2010, IEEE ICCT 2011, ITST 2012, IEEE VTC 2013 Spring, IWCMC 2015, IWCMC 2016, IEEE/CIC ICCC 2016, WPMC 2016, WOCC 2019, IWCMC 2020, WCSP 2020, CSPS2021, WCSP 2021, and IEEE/CIC ICCC 2022.
	
\end{IEEEbiography}

\begin{IEEEbiography}[{\includegraphics[width=1in,height=1.25in,clip,keepaspectratio]{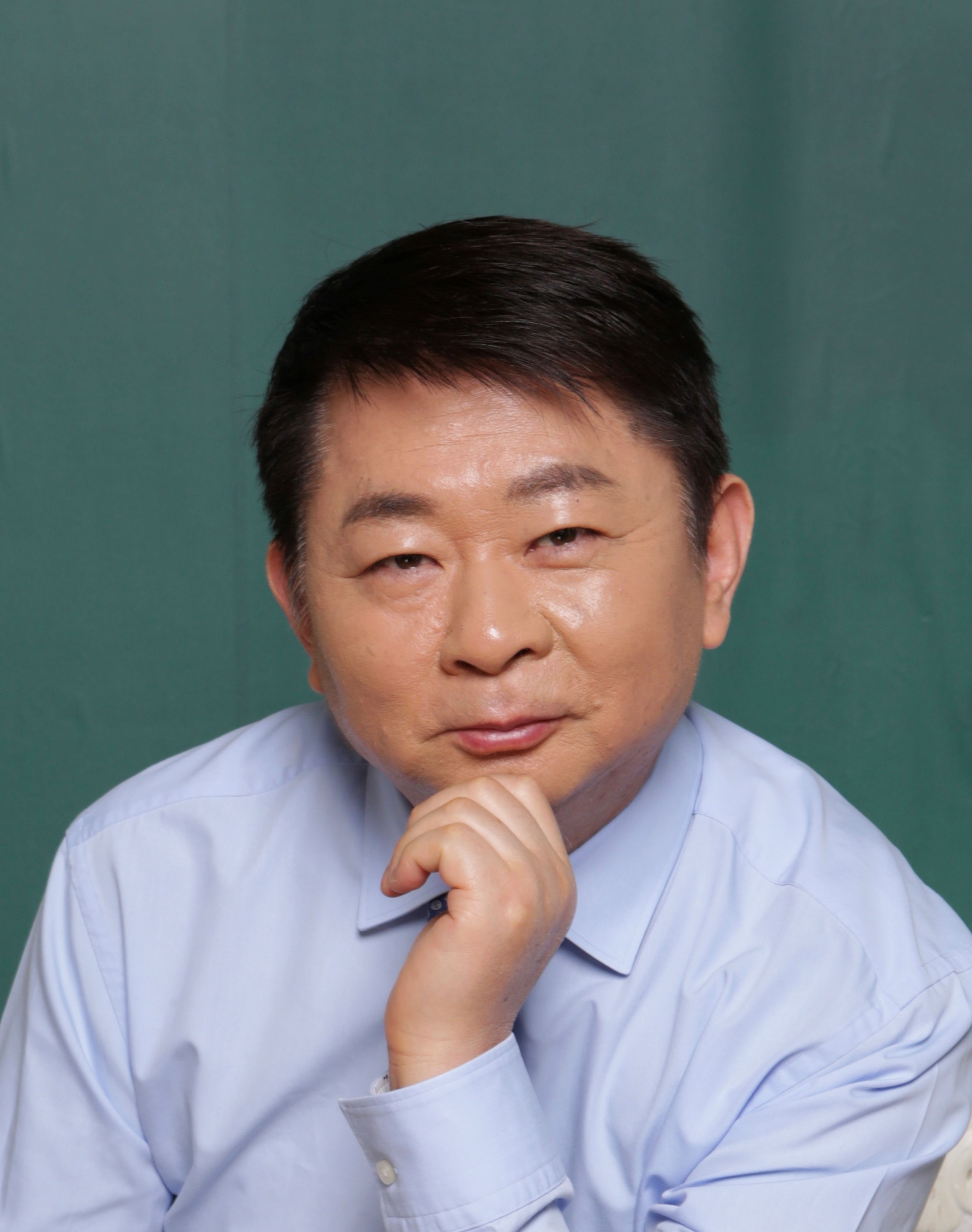}}]{Geoffrey Ye Li}
	(Fellow, IEEE) is currently a Chair Professor at Imperial College London, UK.  Before joining Imperial in 2020, he was a Professor at Georgia Institute of Technology, USA, for 20 years and a Principal Technical Staff Member with AT\&T Labs – Research (previous Bell Labs) in New Jersey, USA, for five years. He made fundamental contributions to orthogonal frequency division multiplexing (OFDM) for wireless communications, established a framework on resource cooperation in wireless networks, and introduced deep learning to communications. In these areas, he has published over 600 journal and conference papers in addition to over 40 granted patents. His publications have been cited over 63,000 times with an H-index of 116. He has been listed as a Highly Cited Researcher by Clarivate/Web of Science almost every year.
	
	Dr. Geoffrey Ye Li was elected to IEEE Fellow and IET Fellow for his contributions to signal processing for wireless communications. He won 2024 IEEE Eric E. Sumner Award and several awards from IEEE Signal Processing, Vehicular Technology, and Communications Societies, including 2019 IEEE ComSoc Edwin Howard Armstrong Achievement Award.
\end{IEEEbiography}

\end{document}